\documentclass[fleqn,10pt]{wlscirep}
\usepackage[utf8]{inputenc}
\usepackage[T1]{fontenc}
\usepackage{amssymb} 
\usepackage{mathtools}
\usepackage{mathdots}
\usepackage{amsmath}
\usepackage{multicol}
\usepackage{braket}
\usepackage{graphicx}
\usepackage{subcaption}
\title{Quasiperiodic disorder induced critical phases in a periodically driven dimerized $p$-wave Kitaev chain}

\author[1,*]{Koustav Roy}
\author[1,2]{Shilpi Roy}
\author[1]{Saurabh Basu}
\affil[1]{Department of Physics, Indian Institute of Technology Guwahati-Guwahati, 781039 Assam, India}
\affil[2]{Department of Physics, National University of Singapore, 117542, Singapore}

\affil[*]{koustav.roy@iitg.ac.in}

\begin{abstract}
The intricate relationship between topology and disorder in non-equilibrium quantum systems presents a captivating avenue for exploring localization phenomenon.  Here, we look for a suitable platform that enables an in-depth investigation of the topic. To this end, we delve into the nuanced analysis of the topological and localization characteristics exhibited by a one-dimensional dimerized Kitaev chain under periodic driving and perform detailed analyses of the Floquet Majorana modes. Such a non-equilibrium scenario is made further interesting by including a spatially varying quasiperiodic potential with a temporally modulated amplitude. Apriori, the motivation is to explore an interplay between dimerization and a quasiperiodic disorder in a topological setting which is also known to demonstrate unique (re-entrant) localization properties. While the topological properties of the driven system confirm the presence of zero and $\pi$ Majorana modes, the phase diagram obtained by constructing a pair of topological invariants ($\mathbb{Z} \times \mathbb{Z} $), also referred to as the real space winding numbers, at different driving frequencies reveal intriguing features that are distinct from the static scenario. In particular, at either low or intermediate frequency regimes, the phase diagram concerning the zero mode involves two distinct phase transitions, one from a topologically trivial to a non-trivial phase, and another from a topological phase to an Anderson localized phase. On the other hand, the study of the Majorana $\pi$ mode unveils the emergence of a unique topological phase, characterized by complete localization of both the bulk and the edge modes, which may be called as the Floquet topological Anderson phase. Moreover, different frequency regimes showcase distinct localization features which can be examined via the localization toolbox, namely, the inverse and the normalized participation ratios. Specifically, the low and high-frequency regimes demonstrate the existence of completely extended and localized phases, respectively. While at intermediate frequencies, we observe the critical (multifractal) phase of the model which is further investigated via a finite-size scaling analysis of the fractal dimension. Finally, to add depth into our study, we have performed a mean level spacing analyses and computed the Hausdorff dimension which yields specific characteristics inherent to the critical phase, offering profound insights into its underlying properties.
\end{abstract}
\begin{document}

\flushbottom
\maketitle
%
%
\thispagestyle{empty}

\section{\label{sec:level1}Introduction}

Anderson localization (AL) is a fundamental phenomenon involving the complete vanishing of transport properties of systems due to the presence of random disorder~ \cite{anderson1,mobilityedge2}. Consequently, all the single-particle eigenstates of the non-interacting system suffer a transition from a completely extended (metallic) to a localized (insulating) phase. While this transition depends upon the dimensionality of the system~\cite{mobilityedge1}, nevertheless, this captivating topic gained paramount interest in a broad range of physical systems, such as matter waves, light waves, optical lattices, photonic lattices, etc~ 
\cite{ALexpt1,ALexpt2,ALexpt3,ALexpt4,ALexpt5}.
On the other hand, quantum systems with incommensurate potential, such as a quasiperiodic (QP) potential, which lies between periodic and random, can exhibit localization transitions in a one-dimensional system \cite{lowdimensionanderson}. Recent developments in experimental accessibility to control over quantum systems and engineering new models in need provide a new era of opportunities for both the experimentalists and theoreticians working in the field. As a consequence, quantum systems with QP potential have been studied in a vast range of experimental setups, including optical \cite{opticallattice1,opticallattice2,opticallattice4} and photonic lattices \cite{photoniclattice1,photoniclattice4}, optical cavities \cite{opticalcavity1,opticalcavity2}, and moir\'{e} lattices \cite{moire}, etc.
To understand the localization transition in QP systems, the Aubry-Andr\'{e} (AA) model \cite{aamodel1,aamodel2} is the most studied one. Later, several general models of the AA model were introduced, comprising of many interesting non-trivial results~\cite{opticallattice4,genaa1,genaa2,genaa3,genaa4,genaa6,genaa7}. 
Moreover, in specific generalized AA models, the transition from an extended to a localized phase is often linked through a critical region, which is characterized as an intermediate or a critical phase \cite{eta}. Numerous studies have shown that an energy-dependent transition that is a mobility edge may be present in this phase~\cite{coexistedge1,coexistedge3,coexistedge4}. The emergence of a mobility edge in a one-dimensional system has sparked substantial curiosity owing to its experimental realizations \cite{mobilityexpt1,mobilityexpt2}. 
\par On a parallel front, topological quantum matter, such as topological insulators (TIs) and topological superconductors (TSCs), has been receiving enormous attention due to its possible applications in topological quantum computation and spintronics devices \cite{spintronics}. The scientific community believes the fundamental building blocks of TSCs are the Majorana zero modes (MZMs), which could be a potential candidate for qubits \cite{qubit1,qubit2}. A simple toy model of the TSCs is provided by the Kitaev chain model~\cite{kitaev}. It is a one-dimensional spinless $p$-wave superconductor chain with MZMs localized at the edges. Various experimental proposals have been projected so far in order to obtain 1D TSCs~\cite{hetero1,hetero2,hetero3,hetero4,hetero5,atomicchain1}. Moreover, several research have been conducted using this model to study the role of QP potential on driving the system towards complete localization~\cite{kitaevIC1,kitaevIC2,kitaevIC3,kitaevIC4,kitaevIC5}. In a similar way, a lot of progress has gained attention from the theoretical aspects. Among them, we are particularly interested in a variant of the generalized Kitaev models. Specifically, a dimerized Kitaev chain \cite{dimerizedkitaev1,dimerizedkitaev2,dimerizedkitaev3}, which is a hybrid of a Su-Schriffer-Heeger (SSH) model \cite{ssh} and a Kitaev chain, owing to its exciting features. There are rich topological and localization properties (protected by the particle-hole symmetry) that make the study of this model intriguing \cite{shilpidimerized}. Furthermore, it is crucial to remember that, as some of us have demonstrated that the interplay between dimerization and the quasiperiodic (QP) potential has been found to have a major impact on a re-entrant localization transition \cite{genaa1}. This led us to consider such a particular choice of the Hamiltonian which involves both the hopping and superconducting pairing being modulated via a dimerization parameter $\delta$. Other versions of this model that also include dimerization in its spin dual representation may be found in the literature, where conventional ferromagnetic and anti-ferromagnetic phases have been studied using several local and non-local order parameters \cite{spinxy1,spinxy2,spinxy3}.
\begin{figure}[t]
\centering
\includegraphics[width=0.75\linewidth]{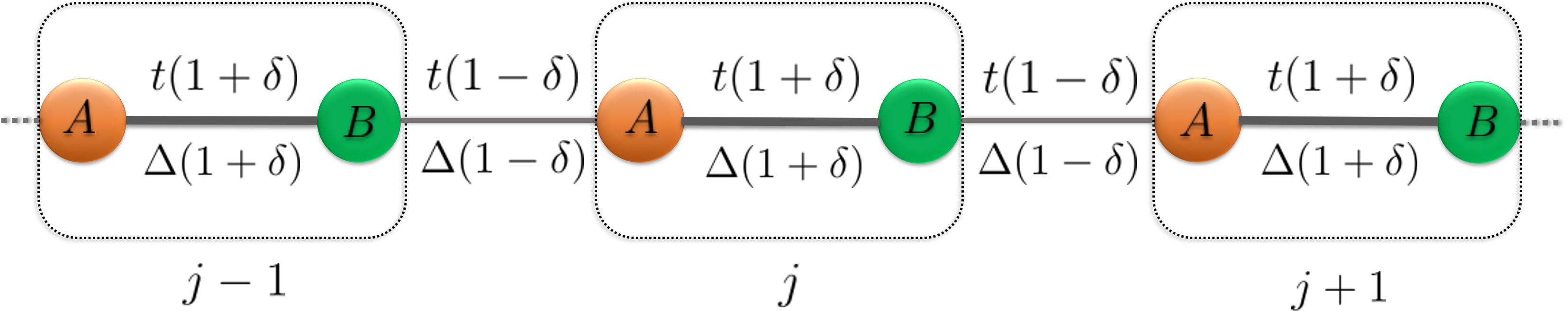}
\caption{Schematic illustration of dimerized Kitaev chain. The red and green circles denote two sublattices, $A$ and $B$ respectively, within a unit cell (dotted line). The thick and thin chain indicates the dimerization of the model. For example, the intracell (strong) and intercell (weak) hopping strengths are defined as $t(1+\delta)$ and $t(1-\delta)$, respectively. Whereas the intracell (strong) and intercell (weak) superconducting pairing strengths are defined as $\Delta(1+\delta)$ and $\Delta(1- \delta)$, respectively.}
\label{Figure_1}
\end{figure}

\par Further, Quantum systems driven periodically far from equilibrium are known to exhibit new phases that are otherwise inaccessible in a static setup. These kinds of periodically driven systems can be understood by means of Floquet formalism \cite{floquetformalism1,floquetformalism2,floquetformalism3}. By using an external periodic drive, one can pave the way for creating topologically non-trivial materials with substantial tunability, even from those that are topologically trivial in the undriven case. Additionally, the energy bands of the driven systems can be backfolded to a Floquet Brillouin zone (FBZ) at the boundary of which, new types of edge modes, namely the so-called $\pi$ modes, appear \cite{floquet1,floquet2,floquet3,floquet4}. Numerous studies using ultracold atoms in driven optical lattices, acoustic, and photonic devices have successfully employed the concept of Floquet engineering \cite{ultracoldfloquet1,ultracoldfloquet2,ultracoldfloquet3,ultracoldfloquet4,ultracoldfloquet5,ultracoldfloquet6,ultracoldfloquet7}. Among others, silicon-on-insulator-like materials with lattices of tightly linked octagonal resonators have been used in the experimental realization of Floquet topological insulators (FTIs) on nano-photonics platform \cite{afzal}. Additionally, photo-induced band gaps can be considered to study the temporal periodicity of systems, where these band gaps can
be resolved using a method called the time and angle-resolved photoemission spectroscopy (t-ARPES) \cite{wangarpes}. In recent years, there has been a remarkable surge in interest revealed to the study of non-equilibrium dynamics of closed systems. These include generation of higher winding or the Chern numbers in 1D, quasi-1D, and 2D systems \cite{rashba,1dfloquet1,1dfloquet2,creutzfloquet,creutzfloquet2,chern1,chern2,paolo1,paolo2}, emergence of time crystalline phases along with period doubling oscillations \cite{period2t1,period2t2}, Floquet analysis of higher-order topological insulators \cite{hoti1,hoti3}, Floquet topological characterization of quantum chaos model \cite{dkr,khm}, etc. In this context, the periodically driven Kitaev chain has been explored in several studies \cite{kitaevfloquet1,kitaevfloquet2}. In addition to that, several studies have been reported on the periodically driven AA model \cite{kickedaa1,kickedaa2,kickedaa3,kickedaa4}.
\par Very recently, the interplay of QP potential and topology has been well explored using this model \cite{shilpidimerized}, where different phases are characterized in terms of topologically trivial and non-trivial regimes. 
In addition to this, the presence of an entire region comprising of multifractal states makes the model more promising in the context of non-ergodic physics \cite{anderson2,ivan,nonergodic}. Moreover, in the presence of random disorder, the model can show behavior similar to the Anderson model \cite{andersondimerized}. Further, a recent study \cite{prb99} has shown that the Majorana modes created using spatially QP driving are more robust to decoherence than those created using a spatially uniform one. 
Deriving motivations from the above inputs, an interesting proposal arose here to bridge the two important aspects, namely the topological properties and the localization-delocalization transition for a time-periodic system, which has been considered by us. Thus, we want to ask very specific questions, such as, how the periodically modulated QP potential affects the above-mentioned properties of the model. Our primary goal here is to compare and contrast the topological and localization properties of a periodically kicked dimerized Kitaev chain corresponding to different frequency regimes. In particular, we observe intriguing nontrivial behavior in the driving setting, which is not present in the static scenario. While the high-frequency regime captures the properties of the static counterpart, the low-frequency regime demands great attention to study deeply. 
\par The layout of the subsequent discussions is as follows. In sec. \ref{sec:level2}, we describe the static version of the model to recapitulate its properties, including dimerization in both hopping and the superconducting pairing term, and benchmark against the results for the driven case. Afterwards, we shall introduce the Floquet tool to construct an effective time-independent Hamiltonian. In sec. \ref{results}, we shall discuss our results on both the topological and the localization properties of the system. Specifically, we have included detailed discussions on the fractal and Hausdorff dimensions and the mean level spacing to distinguish between the localized, critical and extended states. At the end, we summarize and conclude our findings in sec. \ref{conclusions}.

\begin{figure}[t]
\hfill
    \begin{minipage}[t]{0.45\textwidth}
        \centering
        \includegraphics[width=\linewidth]{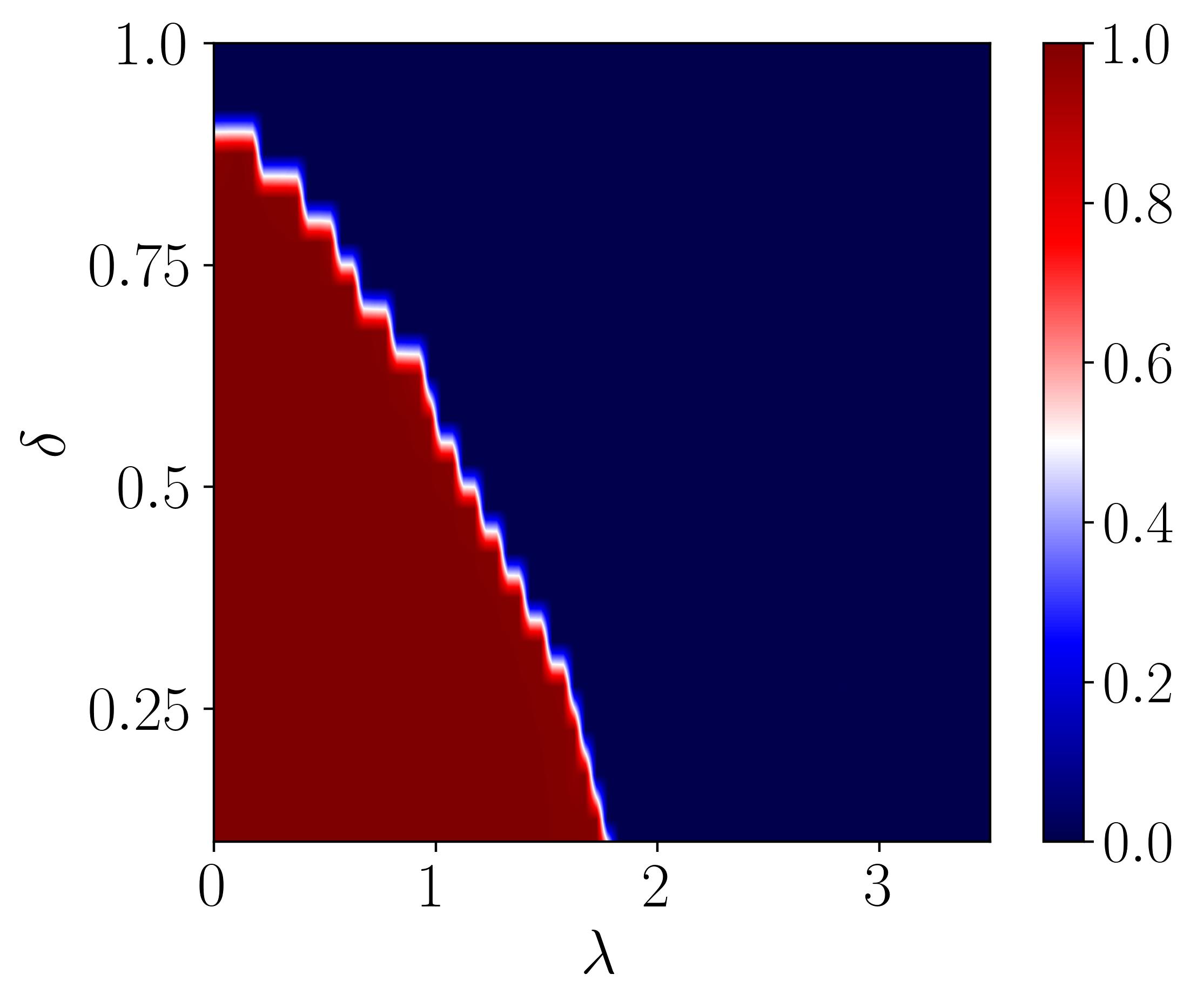}
        \caption{A static phase diagram is shown using the real-space winding number as a function of dimerization strength ($\delta$) and the onsite QP potential strength ($\lambda$). The blue region represents a trivial phase with $\nu=0$, whereas the red region represents a topological phase with $\nu=1$.}
        \label{Figure_2}
    \end{minipage}
    \hfill
    \begin{minipage}[t]{0.45\textwidth}
        \centering
        \includegraphics[width=\linewidth]{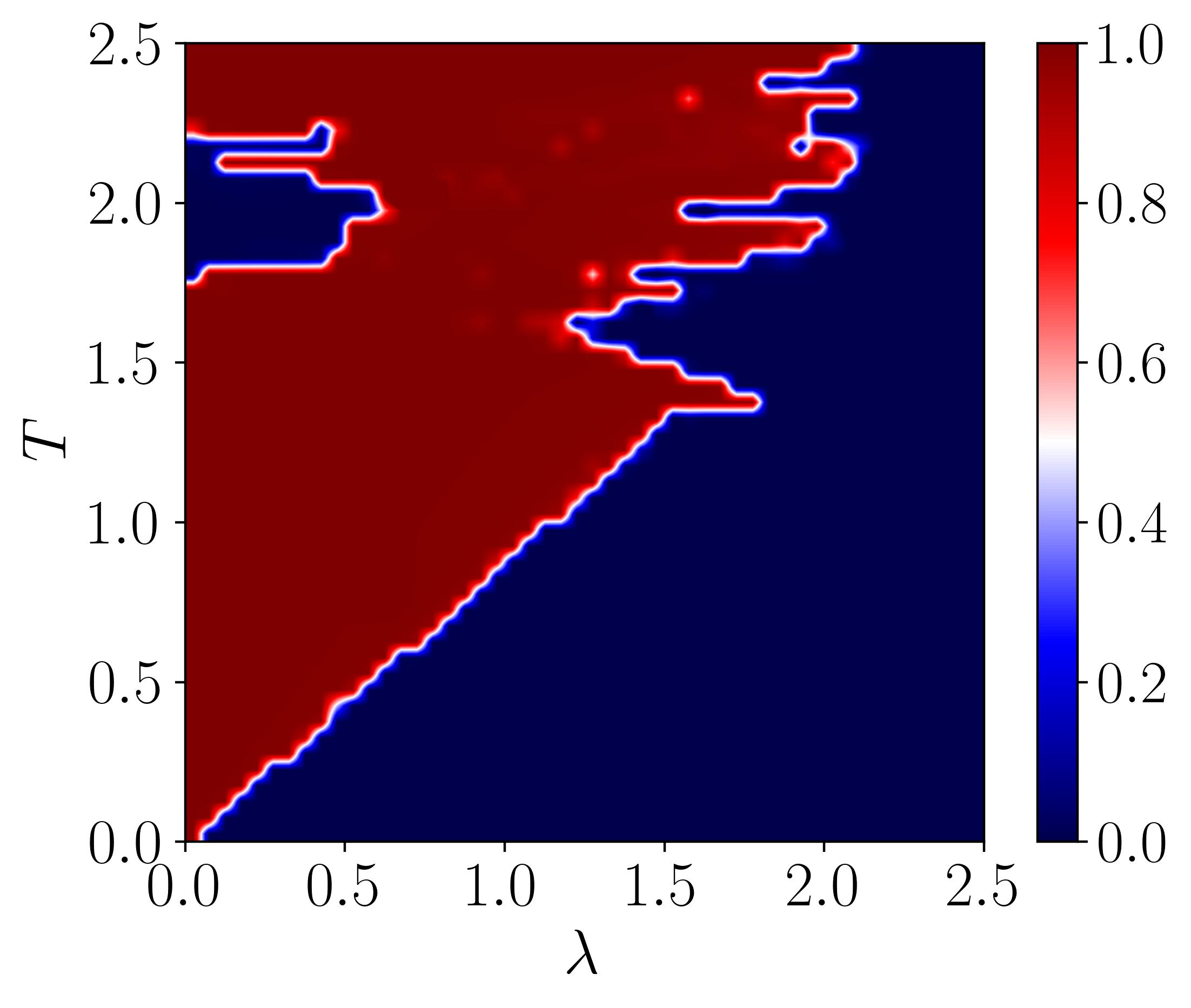}
        \caption{The real-space winding number corresponding to the Majorana-zero modes as a function of the driving period ($T$) and the onsite QP potential strength ($\lambda$) is shown. Here, the dimerization strength is $\delta=0.5$. The blue region represents a trivial phase with $\nu=0$, whereas the red region represents a topological phase with $\nu=1$.}
        \label{Figure_3}
    \end{minipage}
\end{figure}
\section{\label{sec:level2}The Hamiltonian and the Floquet formalism}
We consider a one-dimensional tight-binding model describing the dimerized Kitaev chain of spinless particles with $p$-wave superconductivity in the presence of onsite QP potential, illustrated in Fig.~\ref{Figure_1}. The corresponding Hamiltonian is given by,
\begin{equation}\label{Eqn:Ham}
\begin{split}
    H & =  -t \sum_{j=1}^{N} \Big[ (1+\delta)\hat{c}^{\dagger}_{j,B}\hat{c}_{j,A} +(1-\delta)\hat{c}^{\dagger}_{j+1,A}\hat{c}_{j,B}  + H.c. \Big]  -\Delta \sum_{j=1}^{N} \Big[ (1+\delta)\hat{c}^{\dagger}_{j,B}\hat{c}^{\dagger}_{j,A} +(1-\delta)\hat{c}^{\dagger}_{j+1,A}\hat{c}^{\dagger}_{j,B}  + H.c. \Big] \\ & \quad - \sum_{j=1}^{N} \mu\Big[ \hat{c}^{\dagger}_{A,j}\hat{c}_{A,j} + \hat{c}^{\dagger}_{B,j}\hat{c}_{B,j}\Big].
\end{split}
\end{equation}
Here, $A$ and $B$ represent sublattice indices. The number of unit cells denoted as $N$ corresponds to the unit cell index $j$ ($j=1,2,3,...N$). Thus, the length of the chain is given by $L=2N$.
The creation (annihilation) operator to create (annihilate) a fermion at the sublattice site ($j,A$) and ($j,B$) is given by $\hat{c}^{\dagger}_{j,A}$ ($\hat{c}_{j,A}$) and $\hat{c}^{\dagger}_{j,B}$ ($\hat{c}_{j,B}$), respectively. Further, $t$ is the nearest-neighbor hopping integral, and $\Delta$ denotes the nearest-neighbor superconducting pairing term, which is taken to be real without any loss of generality. It is assumed that the strengths of the hopping integral ($t$) and the $p$-wave superconducting pairing term ($\Delta$) alternate between strong (inside the unit cell) and weak (between the unit cells) bonds. Consequently, a dimerization tuning parameter $\delta$ is introduced in the model to distinguish between them. Hence, the intra (inter) cell hopping integral and the superconducting pairing term are represented by $t(1+\delta)$ $(t(1-\delta))$ and $\Delta(1+\delta)$ $(\Delta(1-\delta))$, respectively, as shown in Fig.~\ref{Figure_1}. We enforce a restriction on $\delta$ to ensure the hopping terms to assume only positive values, namely, $|\delta|<1$. \par Further, in addition to the onsite term, we modulate the chemical potential at sublattice $A$ ($\mu_A$) and sublattice $B$ ($\mu_{B}$) quasiperiodically, given as,
\begin{subequations}
\begin{align}
    \mu_A = \lambda_A \cos{[2\pi \beta (2j-1) + \phi]},
    \\ \mu_B = \lambda_B \cos{[2\pi \beta (2j) + \phi]}.
\end{align}
\label{Eqn:modulation}
\end{subequations}
The periodicity of the QP potential is described by $1/\beta$, where $\beta$ is an irrational number usually chosen to be the golden ratio, namely, $\beta = \frac{\sqrt{5}-1}{2}$. The phase term of the potential is represented as $\phi$, which is set to be zero. The potential strengths at the two sublattice sites are denoted by $\lambda_A$ and  $\lambda_B$. In this paper, we shall mainly focus on the staggered case, namely, $\lambda_A=-\lambda_B=\lambda$. The staggered case is particularly interesting owing to the possibility of an interplay between the dimerization parameter and strength of the QP potential, etc., on the localization properties \cite{genaa1}.
By considering the limiting cases, the dimerized Kitaev chain Hamiltonian is reduced to the Kitaev chain corresponding to $\delta=0$ and to the SSH chain for $\mu=0$ and $\Delta=0$. Throughout this paper, we have fixed $t$ as a unit of energy, that is, $t=1$, and the length of the chain is set as $L=1220$. Additionally, the superconducting pairing strength is set as $|\Delta|<t$, that is, $\Delta=0.5$.

\begin{figure}[t]
\centering
\includegraphics[width=0.7\linewidth]{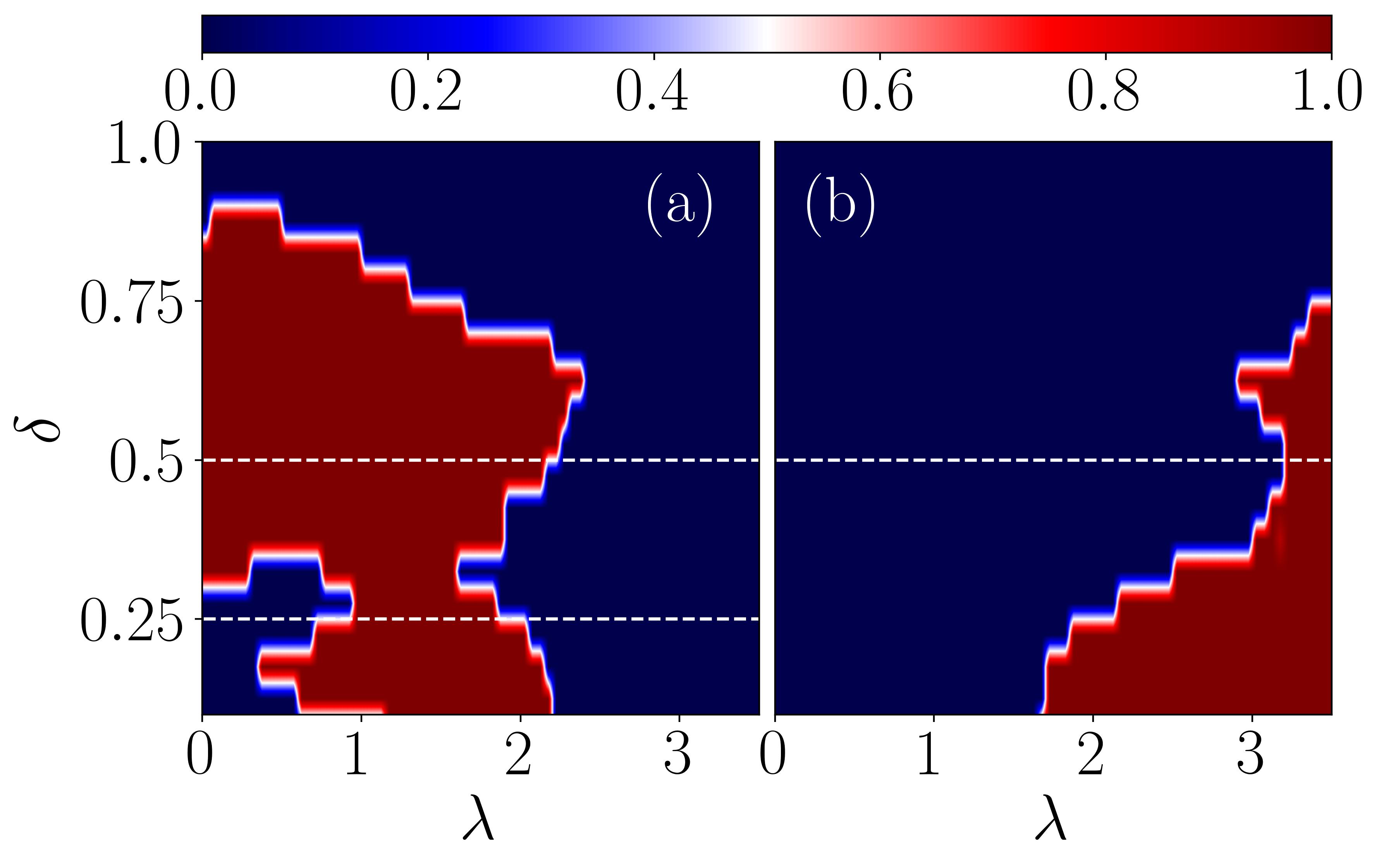}
\caption{The real-space winding number corresponding to the zero and $\pi$ modes as a function of dimerization strength ($\delta$) and the onsite QP potential strength ($\lambda$) are shown in (a) and (b), respectively. Here, the driving frequency is $\omega=2.5$. The system size taken for the calculation is $L=1220$. The blue region represents a trivial phase with $\nu^{0,\pi}=0$, whereas the red region represents a topological phase with $\nu^{0,\pi}=1$.}
\label{Figure_4}
\end{figure}
\begin{figure}[t]
\centerline{\hfill
\includegraphics[width=0.37\textwidth]{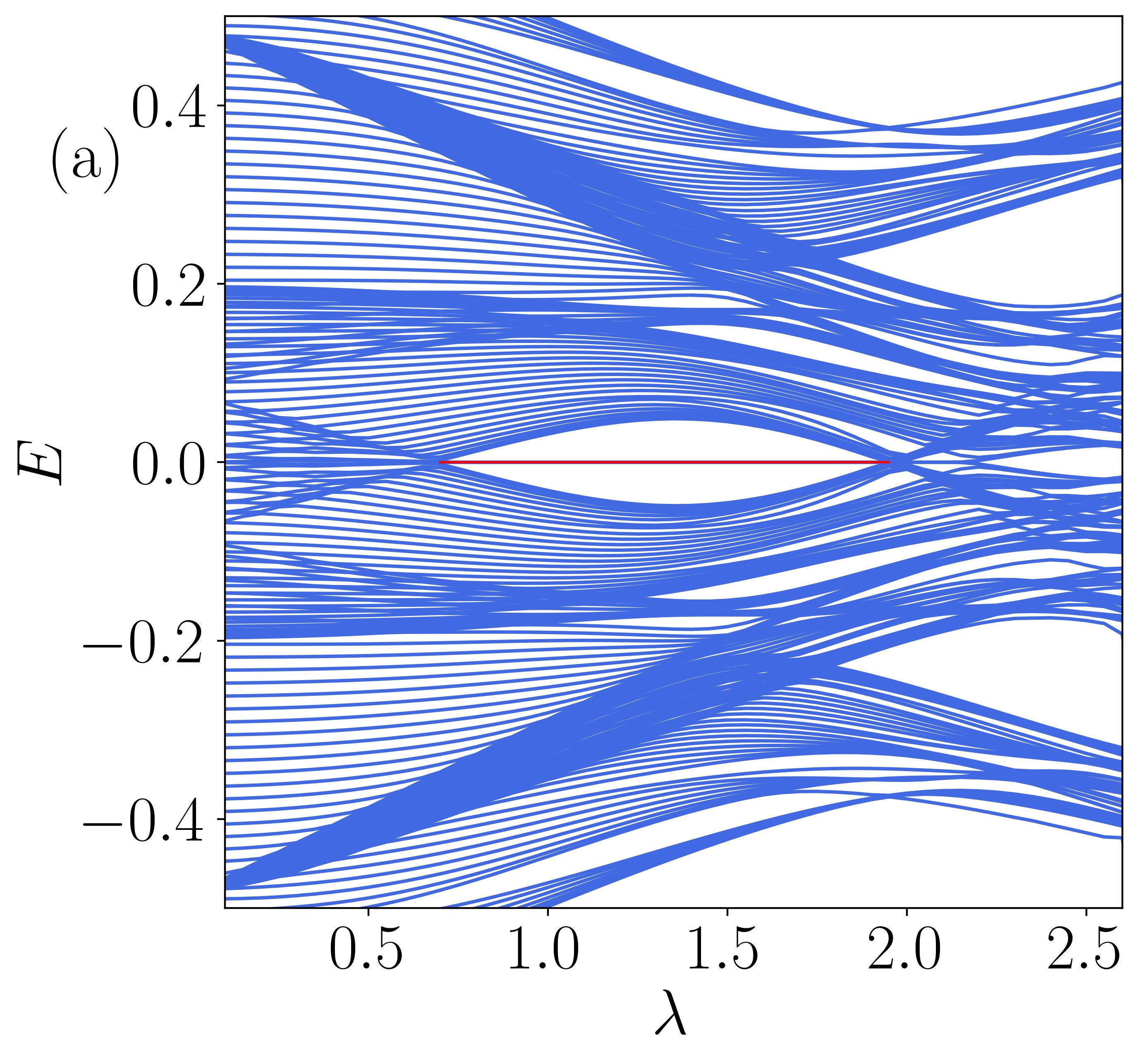}
\includegraphics[width=0.37\textwidth]{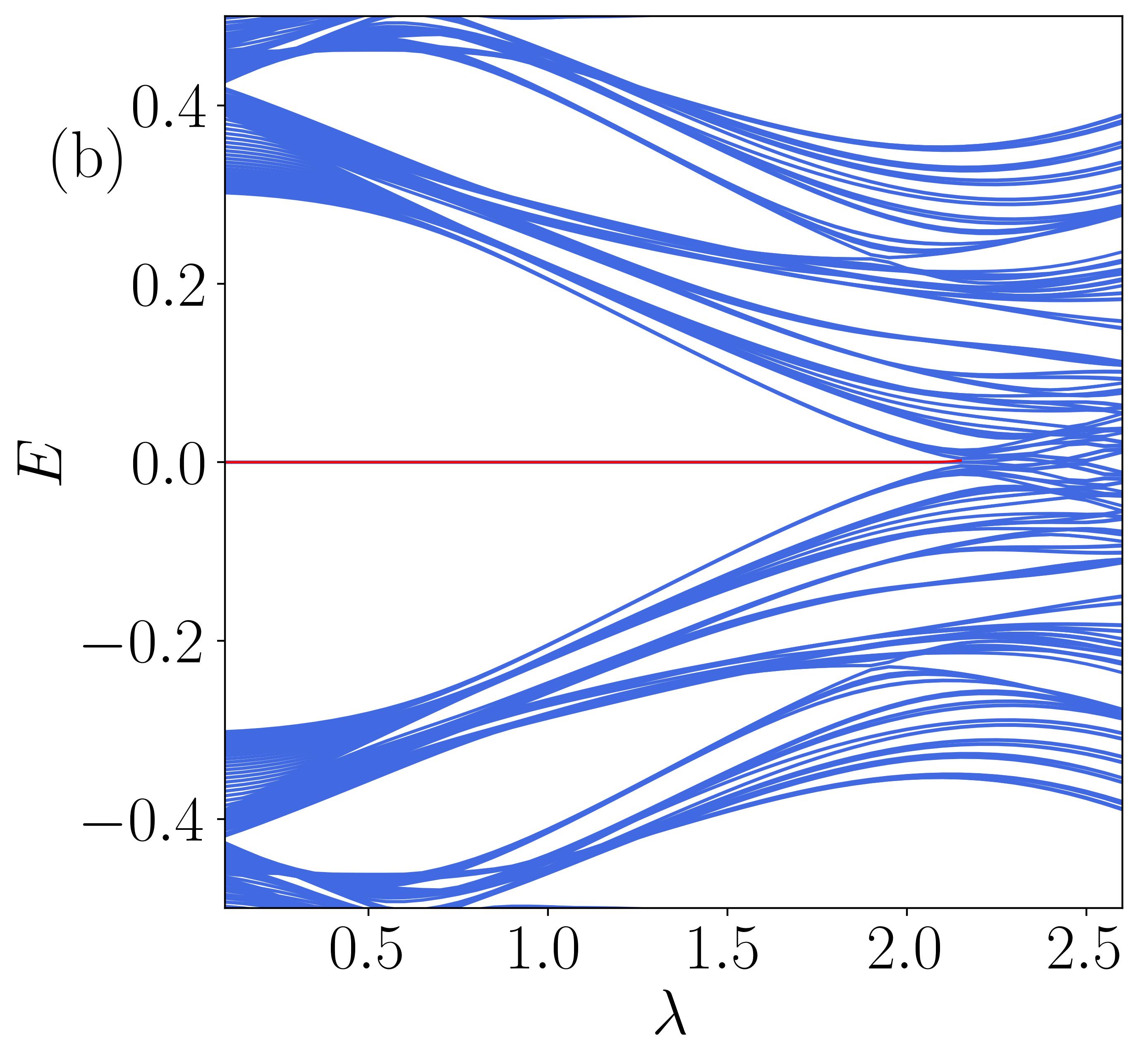}
\hfill
}
\caption{ The Floquet quasi-energy spectrum for a particular frequency, say, $\omega = 2.5$ is shown as a function of $\lambda$ corresponding to (a) $\delta=0.25$ and (b) $\delta=0.5$.}
\label{Figure_5}
\end{figure}
\par Now, to elucidate the effects of periodic driving, we introduce temporal modulations to the QP potentials, given as,
\begin{equation}
    \mu_{A,B} \rightarrow  \mu_{A,B}\sum_{m=-\infty}^{\infty} \delta ( t-mT ).
    \label{Eqn:Driving_protocol}
\end{equation}
Here, the driving protocol (Eq.~\ref{Eqn:Driving_protocol}) indicates applying the onsite QP potential at the sublattice sites $A$ and $B$ periodically at times $t=mT$ where $m$ is an integer that counts the number of kicks.
\par In general, to tackle any periodically driven system, one can adopt the Floquet formalism approach. Which provides a tool to construct an effective time-independent Floquet Hamiltonian whose stroboscopic dynamics are equivalent to the static system. Note that, a detailed analysis of the static system in the presence of QP potential has been studied by some of us in \cite{shilpidimerized}. According to the Floquet theorem, the dynamical evolution of a periodically kicked system is obtained via a time-ordered product of the Floquet evolution operators. As a result, it can be written as a product of two terms, that is,
\begin{equation}
    \hat{U}(T,0)  = \mathcal{T} \int_{0}^{T} e^{-i H(t) dt} = \hat{U}^{\prime} \hat{U}^{\prime \prime}
    \label{Eqn:Ham_drive}
\end{equation}
where,
\begin{equation}
    \hat{U}^{\prime} = e^{-i [\sum_{j=1}^{N} \mu_A  \hat{c}^{\dagger}_{A,j}\hat{c}_{A,j} +  \mu_B \hat{c}^{\dagger}_{B,j}\hat{c}_{B,j}]}
\end{equation}
and
\begin{equation}
    \hat{U}^{\prime \prime} = e^{-i\hat{H}_0T}.
\end{equation}
Here $H_0$ is the static part of the Hamiltonian. We can now numerically diagonalize $\hat{U}(T,0)$ to obtain its eigenvectors $\ket{\psi_m}$ and eigenvalues $e^{-iE_m}$ using,
\begin{equation}
    \hat{U}(T,0)\ket{\psi_m} = e^{-i\hat{H}_{\text{eff}}T} \ket{\psi_m} = e^{-i E_mT} \ket{\psi_m},
\label{effective_Hamiltonian1}
\end{equation}
where $H_{\text{eff}}$ is the Floquet effective Hamiltonian. $E_m$ denotes the quasi-energies which lie within the first FBZ. In the following section, we shall present the exact numerical results obtained by using the Floquet effective Hamiltonian.

\section{\label{results}Results}
Our main aim here is to study the topological and localization properties of the driven system induced by
the interplay of the dimerization strength ($\delta$) and the periodic driving amplitude  ($\lambda$). Hence, based on our numerical analysis, we shall investigate the behavior of the system at different frequencies. In particular, we show different phase transitions and properties of the edge modes and bulk states that are discernible at high- and low-frequency regimes. The system size taken for the numerical calculation is $L=1220$.

\subsection{Topological properties}
In this section, we shall start our discussion on the topological properties based on the study of zero-energy edge modes, that is, MZMs that emerge in our system. Before focusing on our periodically kicked setting, a brief discussion on the static scenario in the presence of the QP potential is useful and presented in the following.
\par The existence of $p$-wave superconductivity in our system implies the particle-hole symmetry is inherently present in the Hamiltonian. Consequently, the quasi-particle energy spectrum is symmetric in nature with respect to the Fermi level ($E_{F}=0$), even in the presence of an onsite QP potential. Thus, for each particle-like solution with energy $+E$, there will be a hole-like solution with energy $-E$. Only the zero-energy states $(E=0)$ are self-conjugated with each other. In this scenario, the topologically non-trivial phase is distinct from the topologically trivial phase by the presence of gapless zero-energy modes. Additionally, this distinction can also be captured via the bulk properties of the Hamiltonian. This yields a topological invariant. Thus, a protected chiral, although with broken translational symmetry (due to the onsite QP potential), hints towards the calculation of the real-space winding number as the topological invariant in our system. In fact, the momentum space formula for the winding number turns out to be useful in this regard. Hence, the real-space winding number ($\nu$) can then be written by drawing an analogy in the momentum space as, \cite{realwinding1,shilpidimerized},
\begin{equation}
    \nu = \frac{1}{L^{\prime}} Tr \Big ( \hat{\Gamma} \hat{Q} \Big [ \hat{Q},\hat{X} \Big]\Big),
    \label{Eqn:real_space_winding}
\end{equation}
where $\hat{Q} = \sum_j^{N} \ket{j}\bra{j} - \ket{\tilde{j}} \bra{\tilde{j}}$ is obtained by solving for a generic chiral symmetric Hamiltonian $H\ket{j} = E_j \ket{j}$ corresponding to $\ket{\tilde{j}} = \hat{\Gamma}^{-1} \ket{j}$, where $j$ is the eigenstate index. $\hat{\Gamma}$ and $\hat{X}$ are operators corresponding to the chiral symmetry and position, respectively. Note that, in the momentum space, the chiral symmetry operators are defined as $\hat{C}= \hat{\sigma}_{0} \otimes \hat{\sigma}_z$ and $\hat{C}= \hat{\sigma}_{x} \otimes \hat{\sigma}_0$ for the cases, corresponding to $\mu=0$ and $\mu\neq 0$, respectively. Hence, the real space representation of the chiral symmetry operator ($\hat{\Gamma}$) can be written as the tensor product of $\hat{C}$ with the corresponding $N^{\text{th}}$ order identity matrix, $I_N$. Finally, $Tr$ denotes the trace over the lattice sites corresponding to half of the length of the chain, namely, $L^{\prime} = L/2$, where half the number of sites are considered from the middle of the chain to eliminate edge effects.
\par The analysis based on the real-space winding number, with the inclusion of a constant chemical potential ($\mu$) in the present scenario results in a distinction between the sublattice and chiral symmetries~\cite{dimerizedkitaev1}. Further, the phase diagram depicting the topological regime via real-space winding numbers in the $\delta-\lambda$ plane is shown in Fig.~\ref{Figure_2}. Subsequently, the onset of the Majorana zero-modes emerges at $\lambda=0$, which is used as a benchmark for making a comparison between the static and the driven scenarios. A disorder-free dimerized Kitaev chain shows a phase transition at $\delta = 0.9$ corresponding to $\mu=1.5$~\cite{dimerizedkitaev2}. On the other hand, introducing the onsite QP potential in the no dimerization limit, that is, $\delta=0$, implies a phase transition from a topologically non-trivial to a trivial at $\lambda<2$. While in the strong dimerization limit ($\delta=1$), the system hosts a trivial phase. As the dimerization strength is increased further from $\delta=0$, it helps in exhibiting the phase transition from the topological to the trivial at the lower value of $\lambda$ with respect to the $\delta=0$ condition.

\begin{figure}[t]
    \centering
    \begin{subfigure}[b]{0.35\textwidth}
        \centering
        \includegraphics[width=\textwidth]{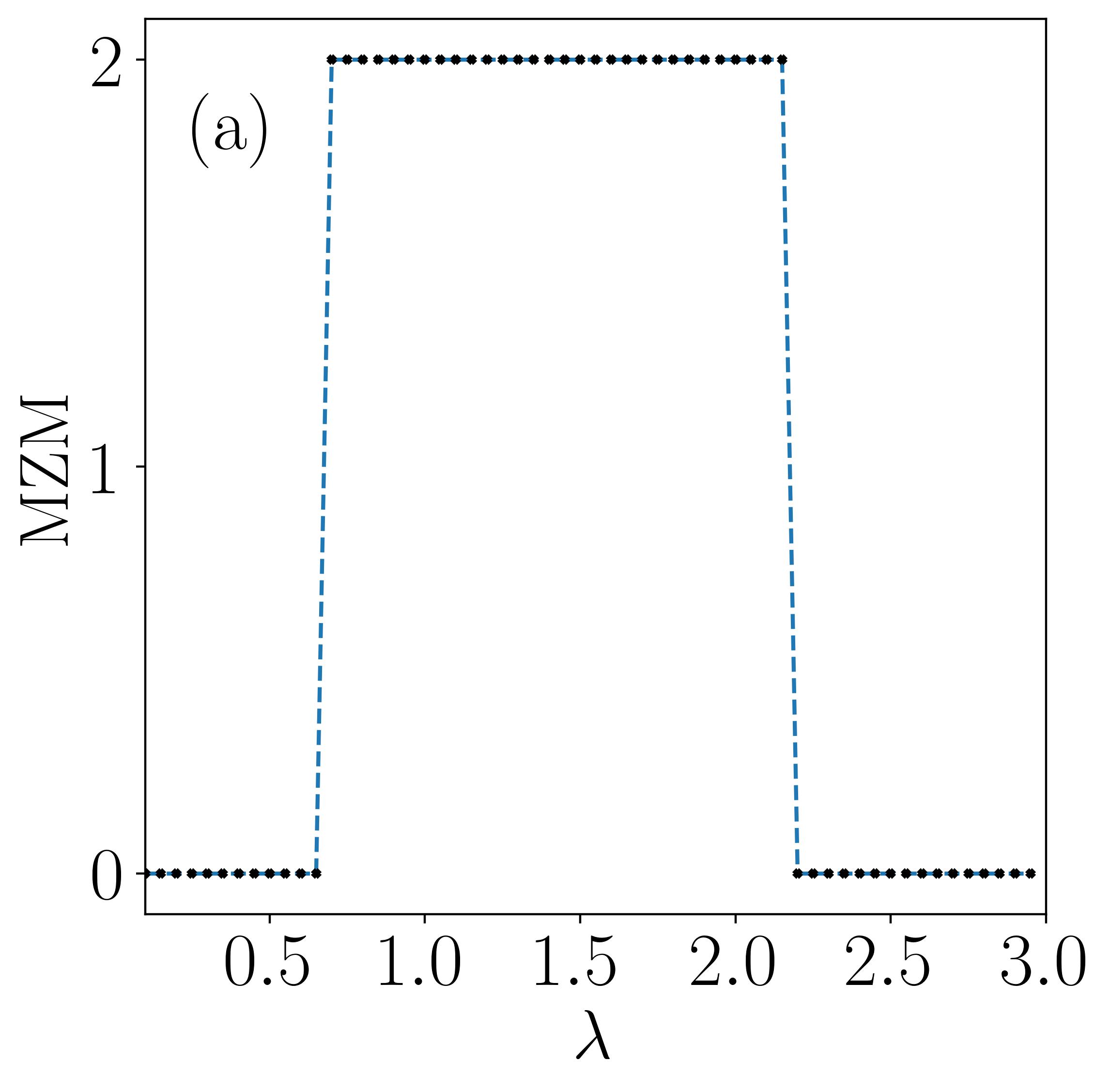}
    \end{subfigure}
    \quad
    \begin{subfigure}[b]{0.35\textwidth}
        \centering
        \includegraphics[width=\textwidth]{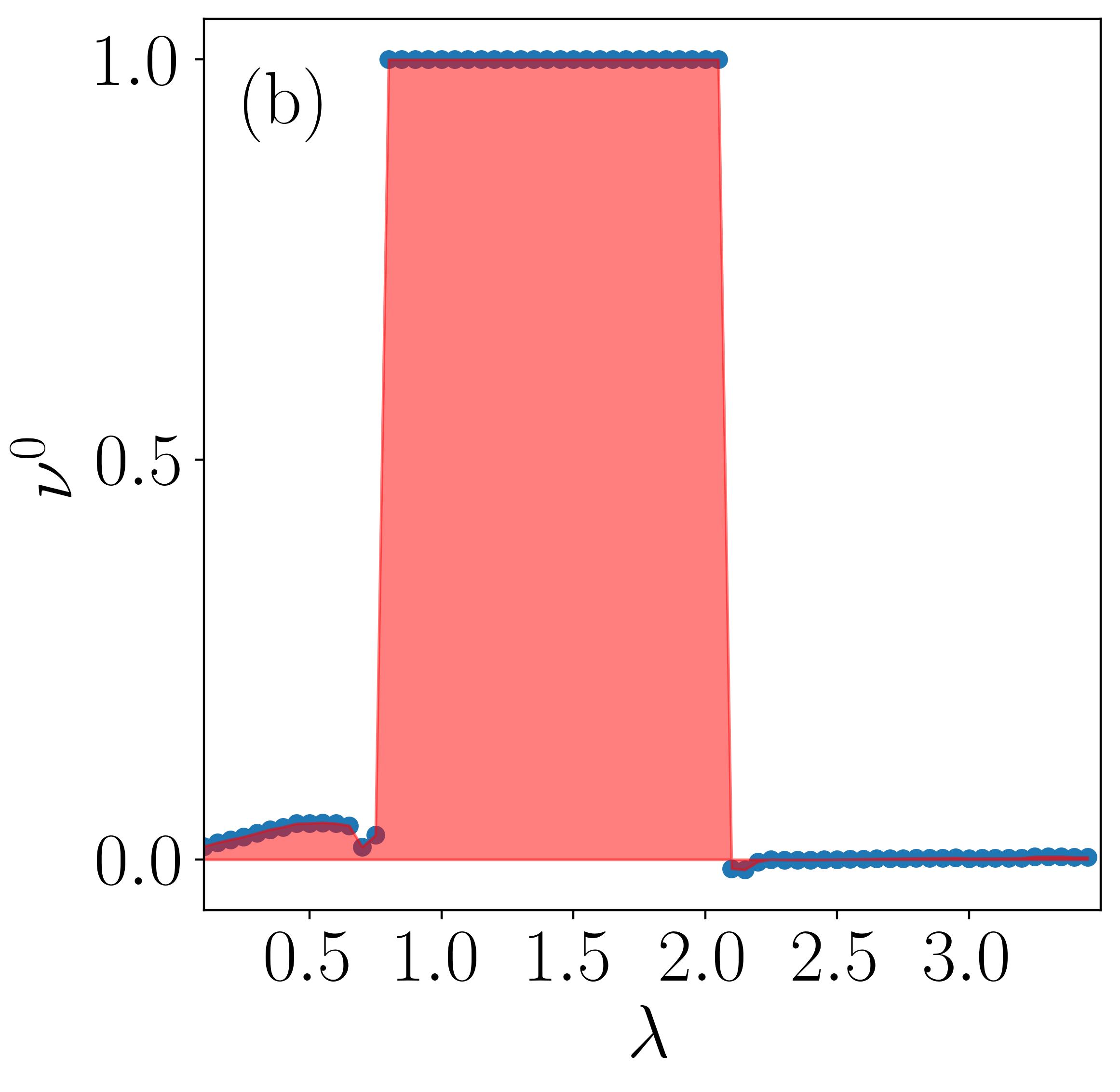}
    \end{subfigure}
    \begin{subfigure}[b]{0.35\textwidth}
        \centering
        \includegraphics[width=\textwidth]{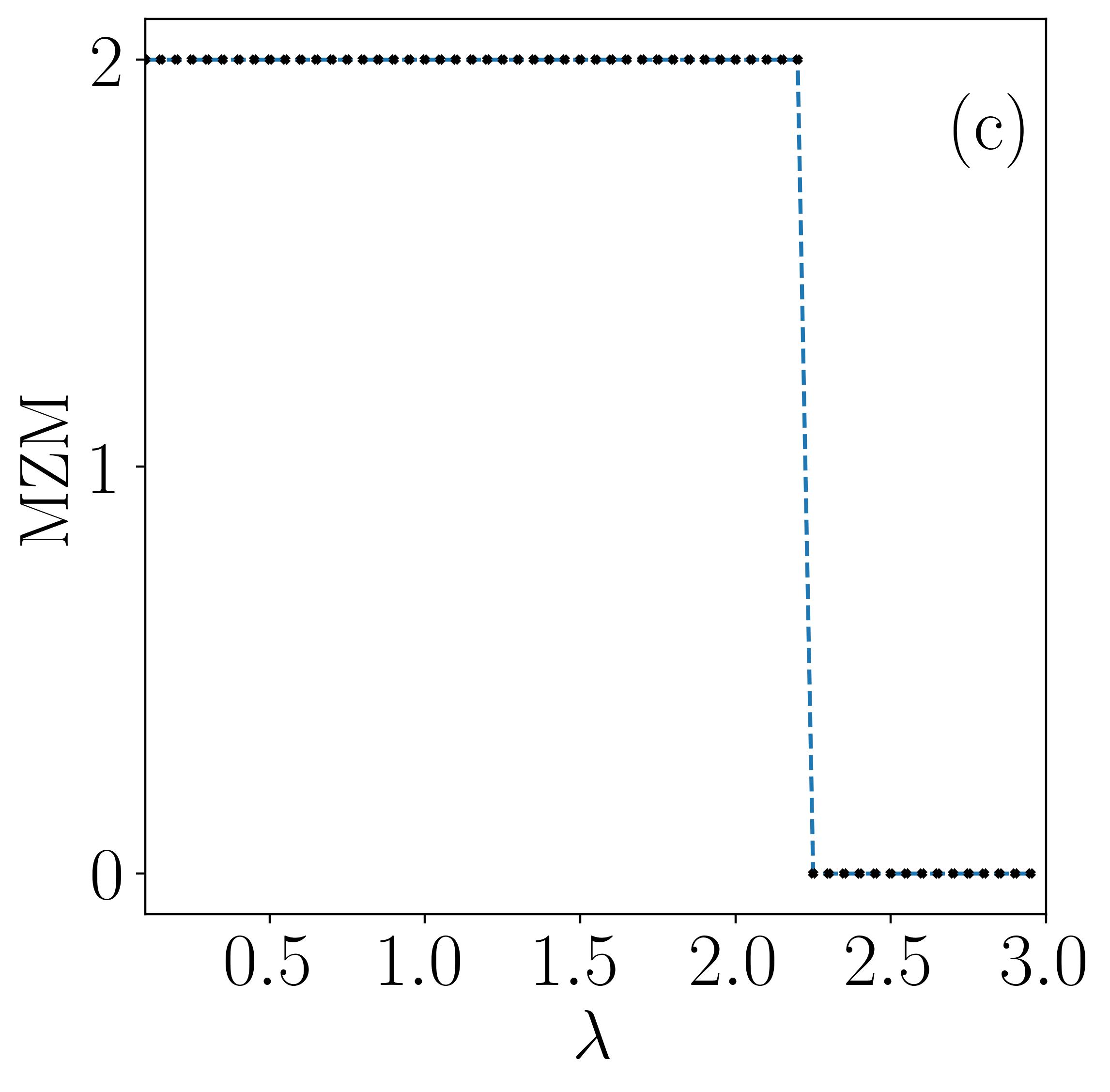}
    \end{subfigure}
    \quad
    \begin{subfigure}[b]{0.35\textwidth}
        \centering
        \includegraphics[width=\textwidth]{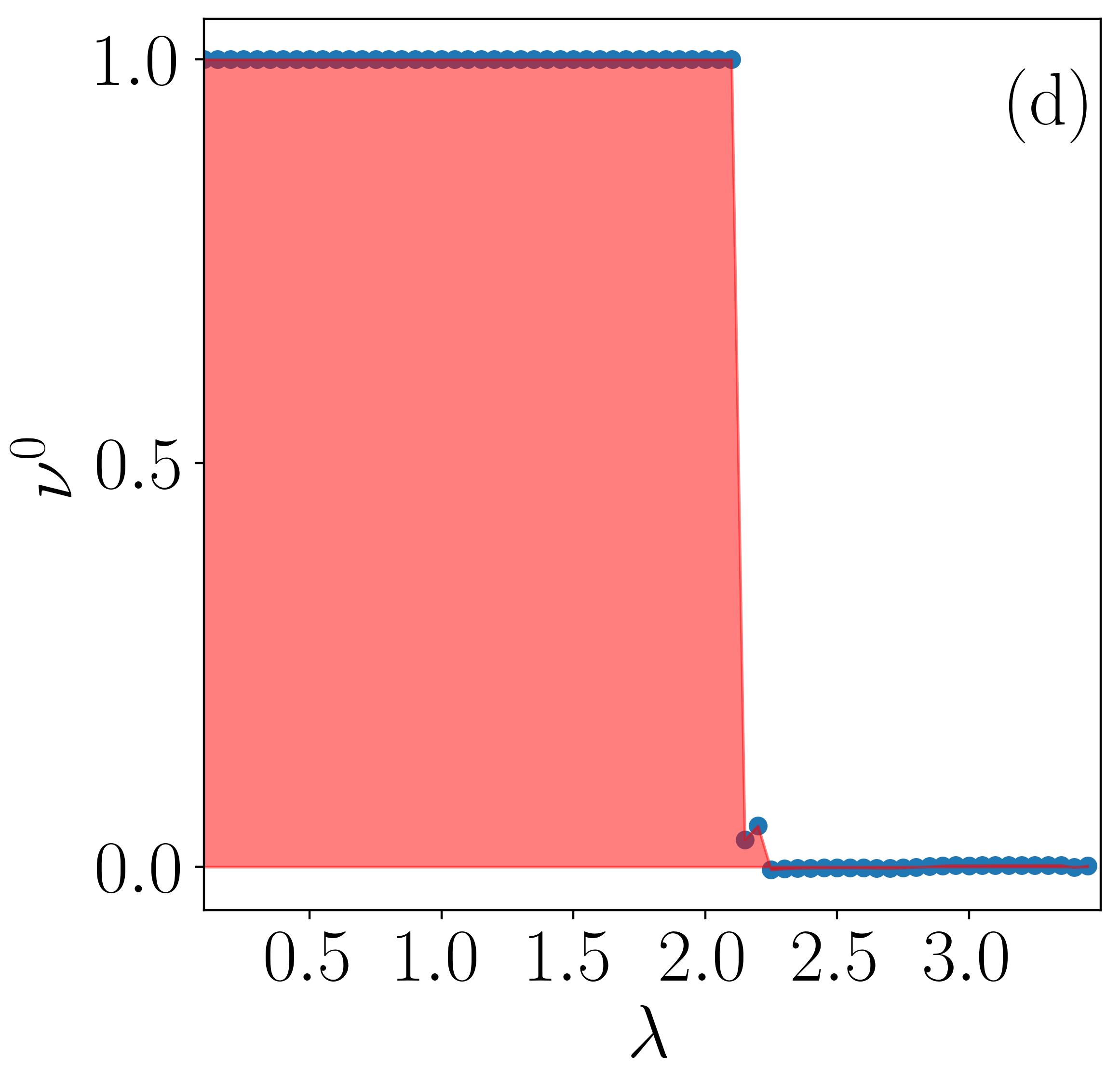}
    \end{subfigure}
    \caption{Majorana Zero modes (MZMs) for a particular frequency, say $\omega = 2.5$ are shown as a function of $\lambda$ in (a) and (c) corresponding to $\delta=0.25$ and $\delta=0.5$, respectively. The real-space winding numbers for zero-energy modes are shown in (b) and (d), corresponding to $\delta=0.25$ and $\delta=0.5$, respectively.}
    \label{Figure_6}
\end{figure}
Expectedly, owing to the unharmed chiral symmetry and periodic drive, the system can host Majorana zero modes. In addition to that, a Majorana $\pi$ mode that has no static counterpart can appear at higher values of the driving strength ($\lambda$). A study of the topological properties in this system is pursued as per the periodic table of FTIs \cite{roy10fold}. Accordingly, it says that each non-trivial phase of the system can be characterized by a pair of winding numbers ($\nu^0,\nu^{\pi}$)  $\in$ ($\mathbb{Z} \times \mathbb{Z} $). The classification of the two non-commutative winding numbers relies on the mechanism of building a pair of symmetric time frames \cite{asbothwinding1,asbothwinding2,creutzfloquet}. Consequently, the Floquet evolution operator acquires a form,
\begin{equation}
    \hat{U} = \hat{F} \hat{G},
    \label{chiral1}
\end{equation}
where, $\hat{F}$ and $\hat{G}$ are related by the chiral symmetry operator ($\hat{C}$) as,
\begin{equation}
    \hat{C} \hat{F} \hat{C} = \hat{G}^{-1}.
    \label{chiral2}
\end{equation}
It is also easy to verify that if a symmetric time frame corresponding to a Floquet evolution operator, $\hat{U}_1 = \hat{F}\hat{G}$
exists, then there must exist another symmetric time frame corresponding to the Floquet operator, $\hat{U}_2=\hat{G}\hat{F}$.
Now the Floquet evolution operator in one symmetric time frame from $t=T/2$ to $t=3T/2$ reads as,
\begin{equation}
    \hat{U}_1 = e^{-i \lambda \hat{V}/2} e^{-i\hat{H}_{0}T} e^{-i \lambda \hat{V}/2},
\end{equation}
where,
\begin{equation}
    \lambda \hat{V} = \sum_{j=1}^{N} [\mu_A \hat{c}^{\dagger}_{A,j}\hat{c}_{A,j} + \mu_B  \hat{c}^{\dagger}_{B,j}\hat{c}_{B,j}].
\end{equation}
\begin{figure}[t]
\centering
\includegraphics[width=0.7\linewidth]{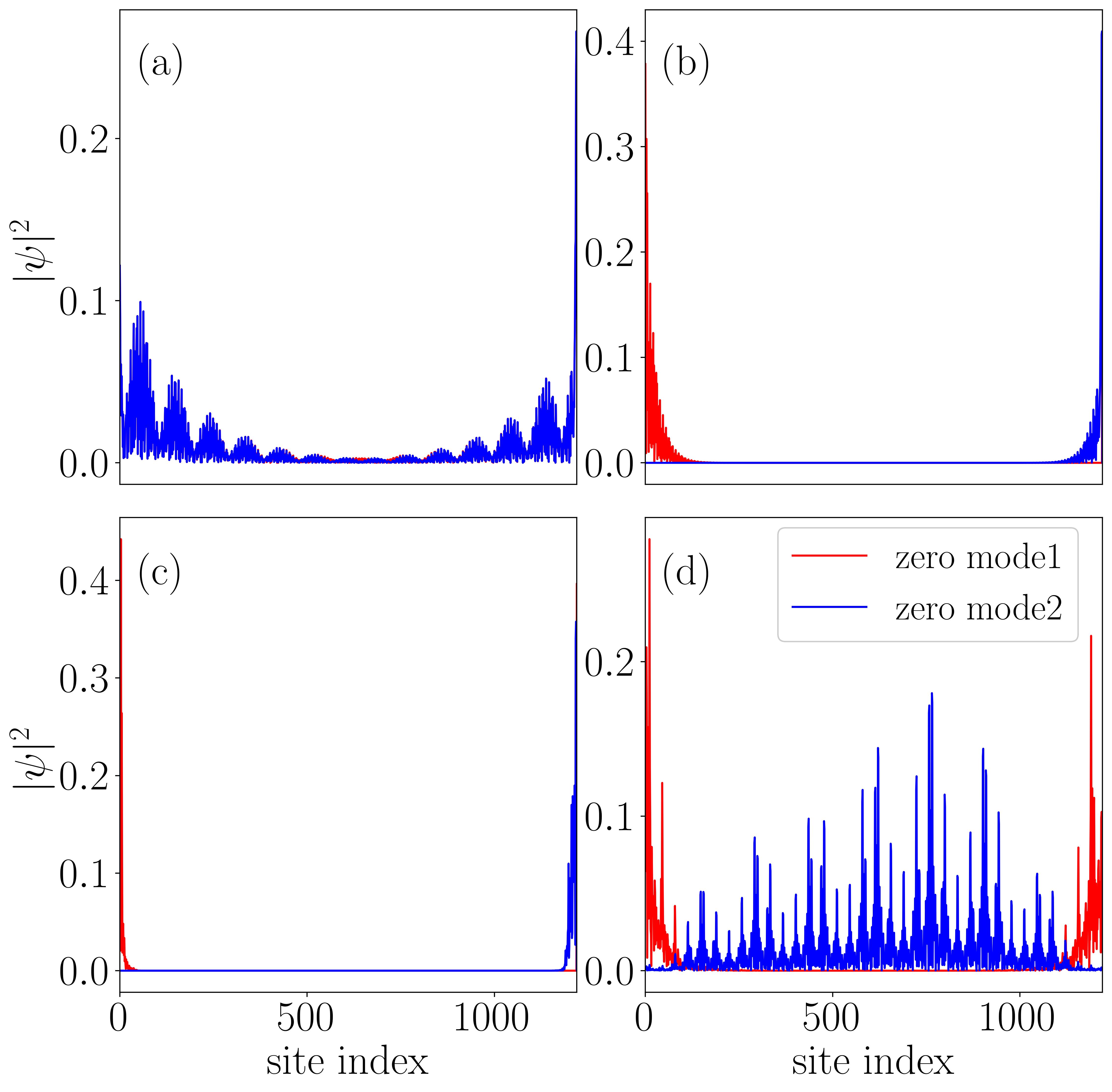}
\caption{The probability distribution of eigenstates as a function of site index $(j)$ are shown corresponding to (a) $\lambda=0.65$, (b) $\lambda=0.7$, (c) $\lambda=1.9$, and (d) $\lambda=2$ for a fixed frequency, say, $\omega=2.5$ and dimerization, say, $\delta=0.25$.}
\label{Figure_7}
\end{figure}
Similarly, using Eq. \ref{chiral1} and Eq. \ref{chiral2}, the Floquet evolution in the second time frame takes the form,
\begin{equation}
    \hat{U}_2 = e^{-i\hat{H}_{0} T /2} e^{ -i \lambda \hat{V}} e^{-i \hat{H}_{0} T/2}.
\end{equation}
Given that $\hat{U}_1$ and $\hat{U}_2$ are chiral symmetric partners and hence share the same quasi-energy spectrum with that of $\hat{U}(T,0)$. Thus, a suitable combination of their winding numbers should be able to provide the topological invariants of the system via the following equations,
\begin{equation}
    \nu^0 = \frac{\nu^{\prime} + \nu^{\prime \prime}}{2} \quad ; \quad \nu^{\pi} = \frac{\nu^{\prime} - \nu^{\prime \prime}}{2}.
\label{invariants}
\end{equation}
Here, $\nu^{\prime}$ and $\nu^{\prime \prime}$ are the winding numbers for the two effective Hamiltonians corresponding to the two symmetric time frames, $\hat{U}_1$ and $\hat{U}_2$ respectively. Meanwhile, we add some pedagogical details by depicting the procedure for obtaining the bulk invariants in a uniformly driven system as outlined in the supplementary material.
\par Now, we start our analysis by enumerating the scenario corresponding to different frequency regimes. Moreover, we want to obtain an upper limit for the frequency which will demarcate between low and high frequency regimes. This can be achieved by plotting the winding number corresponding to the Majorana zero mode ($\nu^{0}$) as a function of the time period $T$ and the driving amplitude $\lambda$ for an arbitrary value of the dimerization strength $\delta=0.5$ in Fig.~\ref{Figure_3}. The phase diagram demonstrates a smooth boundary separating the topological non-trivial phase from a trivial phase in a high-frequency (small period) limit. The observation should emulate the static scenario. On the other hand, in the low-frequency (large period) regime, an intriguing, however highly complex, situation emerges distinct from that of the static counterpart. This non-triviality can be understood by expanding the Hamiltonian using the Baker-Campbell-Hausdorff (BCH) formula given as,
\begin{equation}
\begin{split}
    \text{ln}(e^{X} e^{Y})=  X + Y + \frac{1}{2} [X,Y] + \frac{1}{12}[X-Y,[X,Y]]  - \frac{1}{24} [Y,[X,[X,Y]]] + ...
\end{split}
\label{Eqn:E19}
\end{equation}
Such that the effective Hamiltonian assumes the form,
\begin{equation}
\begin{split}
    \hat{H}_{\text{eff}} =  \hat{H}_0 + \frac{\lambda\hat{V}}{T} + \frac{\lambda T}{2} [\hat{H}_{0},\hat{V}]  + \frac{\lambda T}{12} [  \hat{H}_0, [\hat{H}_0,\hat{V}]] - \frac{\lambda^2}{12} [ \hat{V},[\hat{H}_0,\hat{V}]] + ...
\end{split}
\label{BCH_expansion}
\end{equation}
In the high frequency limit ($T\ll1$) and small driving strength ($\lambda$), $H_{\text{eff}}$ can be truncated upto the first order as,
\begin{equation}
    \hat{H}_{\text{eff}} = \hat{H}_0 + \frac{\lambda\hat{V}}{T}.
    \label{Eqn:renormalized_potential}
\end{equation}
Such an expansion tells us that the Hamiltonian corresponding to a small time period (high frequency) regime is equivalent to the static Hamiltonian plus a renormalized potential (denoted by the second term in Eq.~\ref{Eqn:renormalized_potential}) that increases linearly with the frequency. The inclusion of the second term along with the static Hamiltonian demarcates the different topological phases via a straight line-like boundary in the $\lambda - T$ plane, as shown in Fig.~\ref{Figure_3}. However, in the limit of low-frequency (large time period), one can not ignore the effects of the additional nested commutators in Eq.~\ref{Eqn:E19} that become more important with increasing power of $T$. Clearly, in the low-frequency regime, the drive can induce longer-range interactions. As a result, one can get topologically protected zero-energy modes even where the QP potential strength is high. Furthermore, multiple phase transitions induced by the disorder can be observed in the limit of low frequency. For further study in the low-frequency regime and in order to get insightful results, we fix the value of frequency corresponding to this regime. For example, we take $\omega=2.5$.
\par
\begin{figure}[ht]
\centering
\includegraphics[width=0.7\linewidth]{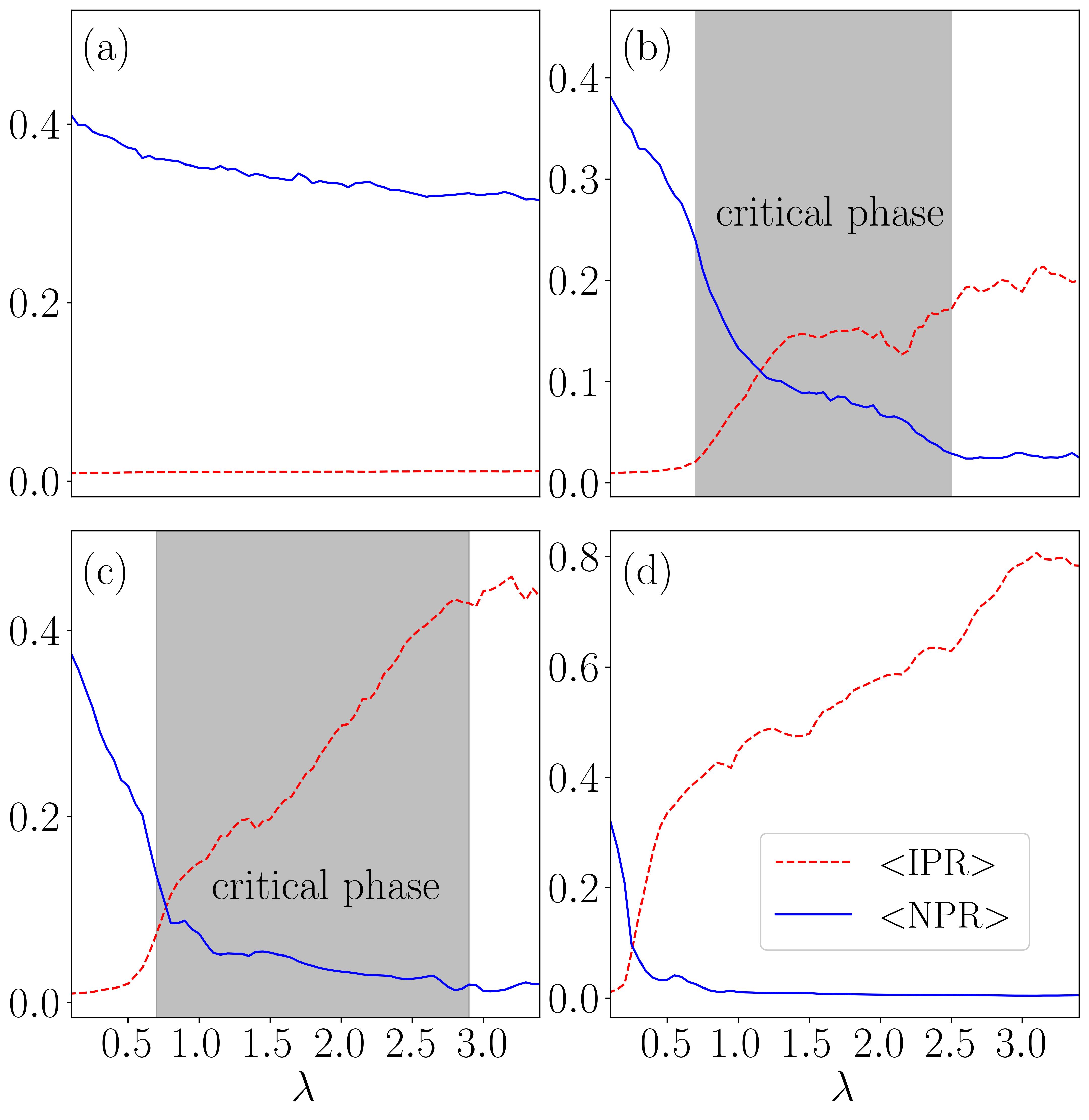}
\caption{The average value of the IPR and the NPR are plotted as a function of $\lambda$ for (a) $\omega=0.5$, (a) $\omega=2.5$, (a) $\omega=3.5$, and (a) $\omega=12.0$. }
\label{Figure_9}
\end{figure}

 We show the phase diagrams via the real space winding numbers ($\nu^{0}$ and $\nu^{\pi}$) in the $\delta-\lambda$ plane corresponding to the above-chosen driving frequency, namely, $\omega=2.5$ in Figs.~\ref{Figure_4} (a) and (b). This value really denotes a representative point in the low-frequency regime, and it hosts a critical phase (See Fig. \ref{Figure_9} and associated discussions below). We shall explain both the phase diagrams and associated analysis in the following.
\par The Majorana zero-energy phase diagram (Fig.~\ref{Figure_4}(a)) of the driven system is seen to host an interesting topological behavior and can be perceived by comparing it with the static scenario (Fig.~\ref{Figure_2}). Specifically, it shows a trivial phase corresponding to weak values for the dimerization strength ($\delta$) and driving amplitude ($\lambda$), which initially was topologically non-trivial in its static counterpart~(Fig.~\ref{Figure_2}).
\begin{figure}[!t]
\centerline{\hfill
\includegraphics[width=0.37\textwidth]{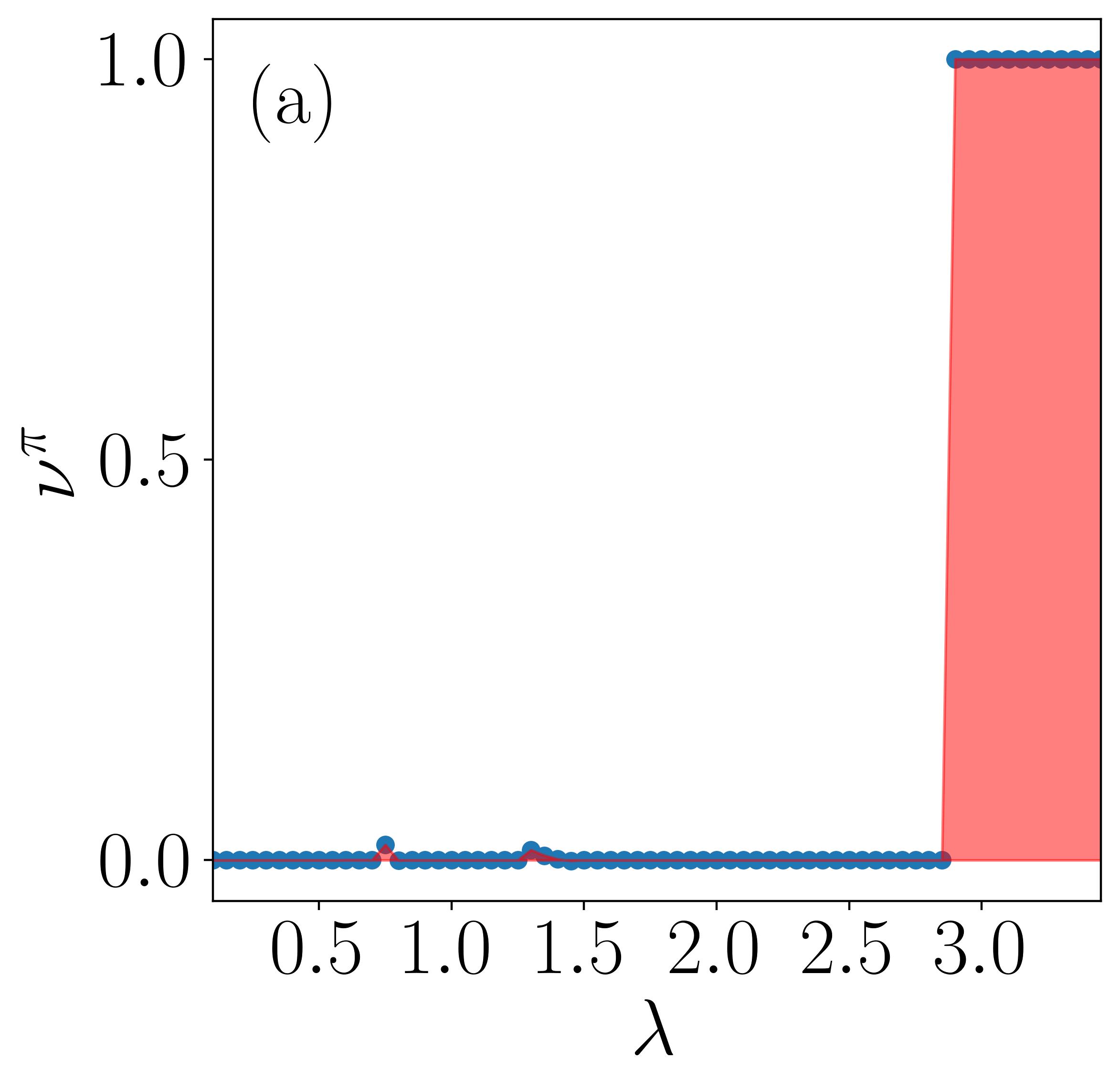}
\includegraphics[width=0.37\textwidth]{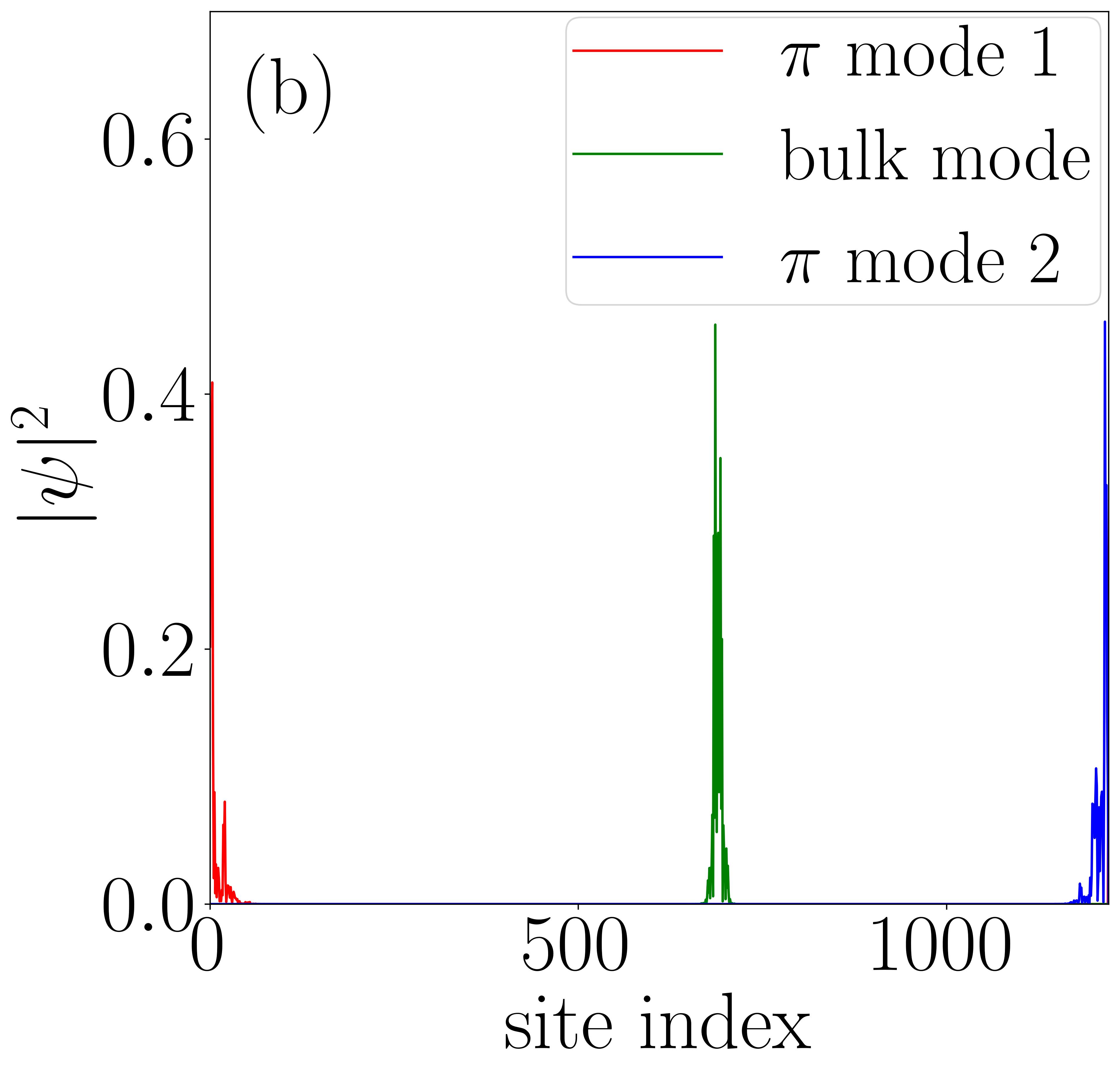}
\hfill
}
\caption{The real-space winding number for $\pi$-energies is shown as a function of $\lambda$ in (a). (b): The probability distributions as a function of the site index are shown corresponding to the $\pi$ modes and the bulk state. The dimerization strength for all the cases is considered as $\delta=0.5$ and the frequency is fixed at $\omega = 2.5$. }
\label{Figure_8}
\end{figure}
Here, we choose two representative dimerization values, namely, $\delta=0.25$ and $0.5$ to further discuss the behavior of the system. In the first case ($\delta=0.25$), the result demarcates a phase transition from a topologically trivial phase to a topologically non-trivial phase at some critical driving amplitude,  
(shown in Fig.~\ref{Figure_4}). Later, upon increasing the driving strength, the system undergoes another phase transition through a second critical point, (shown in Fig.~\ref{Figure_4}) where a transition from a topological non-trivial phase to a trivial Anderson phase occurs. Hence, we may conclude that phase transitions induced by period driving are possible, resulting in the emergence of non-trivial phases, and yields a feature that is unseen in the static scenario (Fig.~\ref{Figure_2}). Further, in the second case ($\delta=0.5$), the topological non-trivial phases expands in the large dimerization regime, resulting in a single-phase transition. Therefore, we observe only a single transition from topologically non-trivial to trivial phases at larger dimerization strengths and perceived only at strong disorder. 
\par In order to acquire a concrete understanding of the topological features observed in the phase diagram, we have shown the Floquet quasi-energy spectrum as a function of $\lambda$ corresponding to two different values of the dimerization strength, say, $\delta=0.25$ and $0.5$ in Figs.~\ref{Figure_5}(a) and \ref{Figure_5}(b), respectively. The choice of $\delta=0.25$ derives motivation from the left panel of Fig.~\ref{Figure_4}, where an initial trivial phase is driven into a topological phase beyond a certain value of $\lambda$, which eventually gets destroyed at larger values of $\lambda$. Thus, a topologically non-trivial phase corresponding to the Majorana zero modes appears in the spectrum, implying phase transitions occurring at two values of $\lambda$, that are, namely, $\lambda_{1} \simeq 0.65$ and $\lambda_{2} \simeq 2.00$. These values are shown in Fig.~\ref{Figure_4}(a) by the intersection of the white dashed line with the red region. On the other hand, corresponding to $\delta=0.5$, we observe the onset of the topological phase from the left edge of Fig.~\ref{Figure_4}, that is, $\lambda=0$, implying the emergence of MZMs. The topological phase exists upto $\lambda \simeq 2.2$, beyond which the MZMs hybridize with the bulk. Subsequently, the number of MZMs and the winding number ($\nu^{0}$) as a function of $\lambda$ confirm the validity of the bulk-edge correspondence, which is shown in Figs.~\ref{Figure_6}(a) and \ref{Figure_6}(b) corresponding to $\delta=0.25$.
In addition to that, we show the same for the other value of $\delta$, namely, $\delta=0.5$ in Figs.~\ref{Figure_6}(c) and \ref{Figure_6}(d). Note that all these observations validate the phase diagram presented in Fig.~\ref{Figure_4}(a).
\par Next, we have plotted the probability distribution of the zero-energy Floquet eigenstates as a function of the site index around the first transition point, $\lambda_{1}$ corresponding to $\lambda=0.6,$ and $0.7$ in Figs.~\ref{Figure_7}(a) and \ref{Figure_7}(b), respectively for $\delta=0.25$. A phase transition supported by the extended and localized nature of the zero-energy edge modes precisely indicates the existence of the first transition around the critical point $\lambda_{1}=0.65$. On the other hand, we have plotted the same corresponding to other values of $\lambda$ that lie on either side of the second transition point; namely, we have taken $\lambda=1.9$ and $2.0$ in Figs.~\ref{Figure_7}(c) and \ref{Figure_7}(d), respectively. The complete localization of the states at the edges validates the signature of the MZMs, implying a topological non-trivial phase being present at $\lambda=1.9$. With an increase in the driving strength, the probability distribution at $\lambda=2.0$ shows a critical behavior at the transition point, thereby indicating a transition to an Anderson phase. 
\par Further, we explore the non-trivial topological features that emerge corresponding to the Majorana $\pi$-mode in the system. Expectedly, it is clear from Fig.~\ref{Figure_4}(b) that the Majorana $\pi$-modes appear at a strong driving amplitude $\lambda$ with respect to the MZMs corresponding to the same values of the parameters as considered for the discussion of the MZMs. Here, we shall study the driving induced features by focusing on a specific value of $\delta$, such as $\delta=0.5$, where we observe a phase transition occurring at $\lambda\simeq 3$ (See Fig. \ref{Figure_4}(b)). For further clarification of the situation, we plot the winding number corresponding to $\pi$-modes in Fig.~\ref{Figure_8}(a). The data clearly indicate a phase transition from the topologically trivial to a topologically non-trivial phase around $\lambda \simeq 3$. Moreover, we observe the probability distribution of the $\pi$-modes and one of the bulk states as a function of the site index in Fig.~\ref{Figure_8}(b). Most interestingly, we find the $\pi$ energies to be localized at the edges along with a completely localized bulk. We have checked that all the bulk states are localized in nature. Therefore, a non-trivial phase comprising of the $\pi$ energy edge modes in addition to a complete localized bulk state, which may be called the Floquet topological Anderson phase, is observed in this system.
\begin{figure}[ht]
\centering
\includegraphics[width=0.7\linewidth]{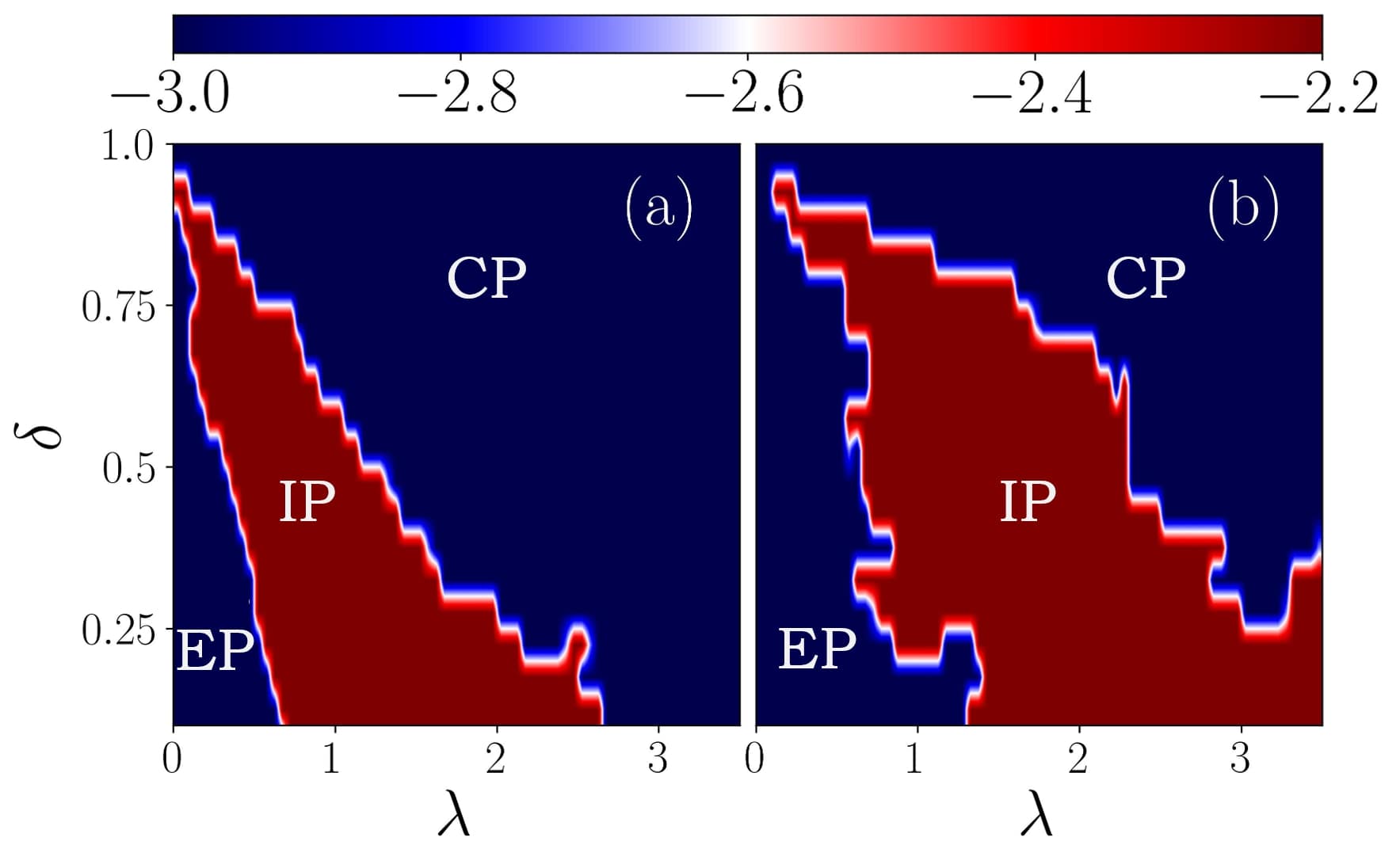}
\caption{A static phase diagram is shown in panel (a) using the parameter $\eta$ as a function of dimerization strength ($\delta$) and the onsite QP potential strength ($\lambda$). Whereas, panel (b) depicts the $\eta$ phase diagram corresponding to a driven scenario ($\omega = 2.5$). Extended, intermediate, and localized phases are shown via `EP', `IP' and `LP' respectively. }
\label{Figure_10}
\end{figure}
\subsection{Localization properties}
In this section, we shall explore the localization properties of bulk states in the periodically kicked setting. To do that, we employ two diagnostic tools, namely, the inverse participation ratio (IPR) and the normalized participation ratio (NPR), to distinguish between the extended, critical, and localized phases.
Similar to the non-driven case, the effective Floquet Hamiltonian can be solved using the Bogoliubov transformation. This can be done by defining a quasiparticle operator in terms of the superposition of the single-particle creation ($c^{\dagger}$) and annihilation ($c$) operators via,
\begin{equation}
    \Phi_n^{\dagger} = \sum_{j=1, \alpha= A,B}^{N} \Big [ u_{j,\alpha}^{(n)} \hat{c}^{\dagger}_{j,\alpha} +  v_{j,\alpha}^{(n)} \hat{c}_{j,\alpha} \Big],
\end{equation}
where $\alpha$ and $n$ denote the sublattice and the energy band indices, respectively. While $u_n$ and $v_n$ denote the particle and hole coefficients. Hence, we can define the IPR and NPR corresponding to the $n^{\text{th}}$ Floquet eigenstate using $u_j^{n}$ and $v_j^{n}$ as, \cite{genaa1}
\begin{equation}
    \text{IPR}^{(n)} = \sum_{j=1, \alpha= A,B}^{N} \Big [ |u_{j,\alpha}^{(n)}| ^4 + |v_{j,\alpha}^{(n)}| ^4 \Big]
    \label{Eqn:IPR}
\end{equation}
and,
\begin{equation}
    \text{NPR}^{(n)} = \Bigg [ L \sum_{j=1, \alpha= A,B}^{N} \Big [ |u_{j,\alpha}^{(n)}| ^4 + |v_{j,\alpha}^{(n)}| ^4 \Big] \Bigg]^{-1}.
    \label{Eqn:NPR}
\end{equation}
In the thermodynamic limit, the IPR(NPR) values scale with the system size as $L^{-1}(L^{0})$ corresponding to an extended state. On the other hand, it varies as $L^{0}(L^{-1})$ for a localized state. Moreover, one can calculate the average values of the IPR and NPR averaged over all the Floquet eigenstates, given by,
\begin{equation}
    \langle\text{IPR}\rangle = \frac{1}{L} \sum_{n=1}^{L} \text{IPR}^{(n)} \quad \text{and} \quad \langle\text{NPR}\rangle = \frac{1}{L} \sum_{n=1}^{L} \text{NPR}^{(n)}.
\end{equation}
In Fig.~\ref{Figure_9} we have shown the variation of $\langle\text{IPR}\rangle$ and $\langle\text{NPR}\rangle$ as a function of $\lambda$ corresponding to different frequencies, such as $\omega=0.5,~2.5,~3.5,$ and $12.0$. We found all the eigenstates to be extended in nature corresponding to the low-frequency limit, that is, for $\omega=0.5$ in Fig.~\ref{Figure_9}(a). Also, all the states are localized for any value of $\lambda$ at a large frequency, namely, $\omega=12$, shown in Fig.~\ref{Figure_9}(d).  
The situation can be understood by the BCH expanded the Hamiltonian given in Eq. \ref{BCH_expansion}.
Now, in the limit of high frequency, the renormalized potential dominates over the rest of the parameters in the static counterpart. As a result, we find that at relatively high (low) driving frequencies, all the states become localized (extended). Whereas, at some intermediate frequencies, such as $\omega=2.5$ and $3.5$, we find the emergence of all three phases, namely extended, critical, or the multifractal, and localized phases with increasing value of $\lambda$. Also, once the system is in the localized phase, it remains localized and is independent of the choice of $\lambda$. However, the multifractal nature of the states can not be uniquely determined via IPR or NPR alone. Additionally, we are interested in the global properties of the model. To this end, we shall introduce another quantity $\eta$, which is defined as \cite{eta,genaa1},
\begin{equation}
    \eta = \text{log}_{10} [ \langle \text{IPR} \rangle \times \langle \text{NPR} \rangle ].
\end{equation}
If either of the phases, such as an extended or a localized phase, is present in the system, the condition $\eta\leq-\text{log}_{10}(L)$ arises from the fact that the value survives owing to the finiteness of either of the average values of IPR 
or NPR. \textcolor{red}{Consequently, an intermediate phase which can be thought of as an admixture of localized and extended states can emerge when both $\langle \text{IPR} \rangle$ and $\langle \text{NPR} \rangle$ are finite. This leads to the condition, $\eta\geq-\text{log}_{10}(L)$. Therefore, in our case, where we have 
considered $L=1220$, $\eta\ge-3.08$ would denote an intermediate phase.}
\begin{figure}[t]
    \centering
    \begin{subfigure}[t]{0.33\textwidth}
        \centering
        \includegraphics[width=\textwidth]{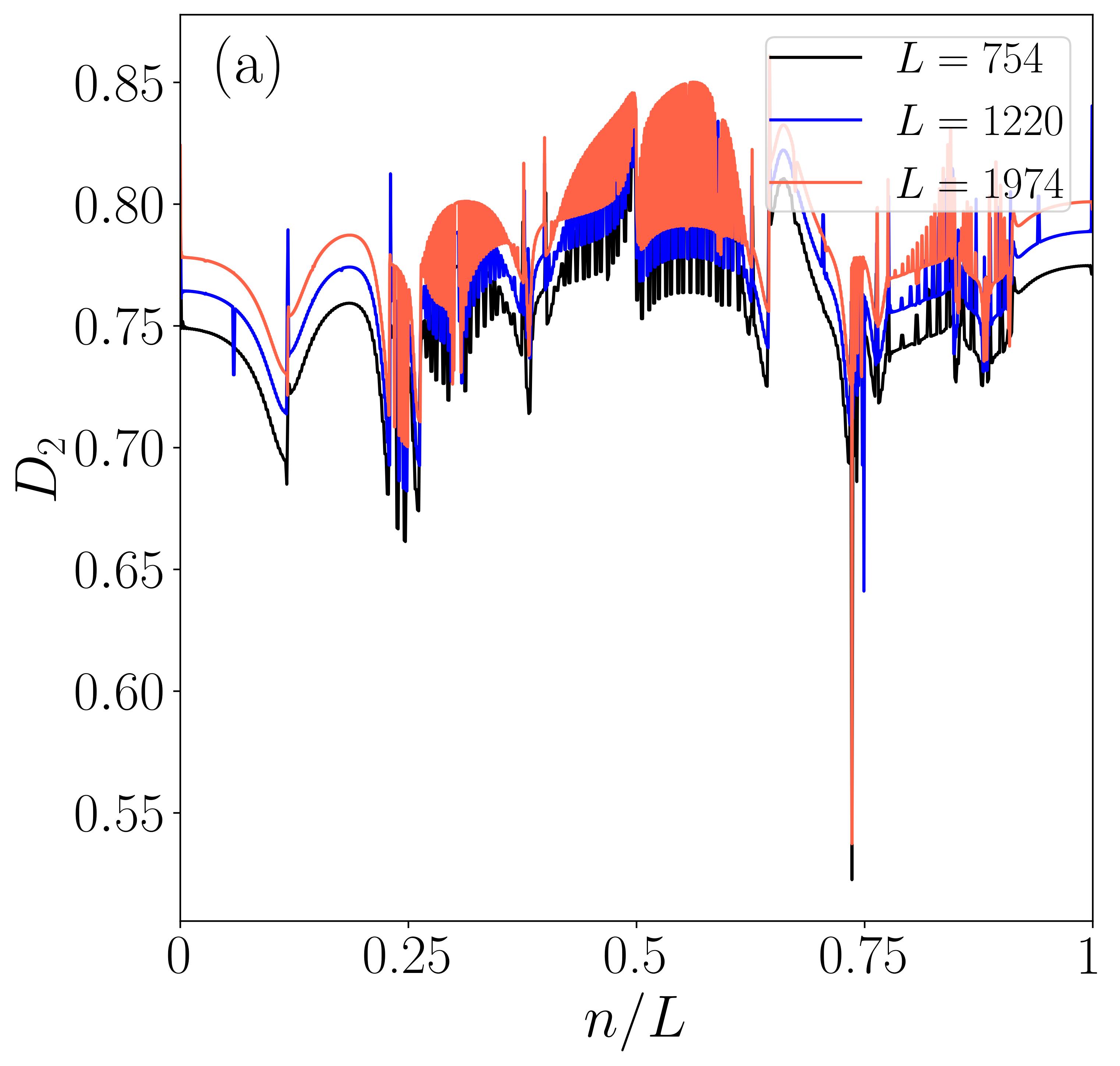}
    \end{subfigure}
    \quad
    \begin{subfigure}[t]{0.32\textwidth}
        \centering
        \includegraphics[width=\textwidth]{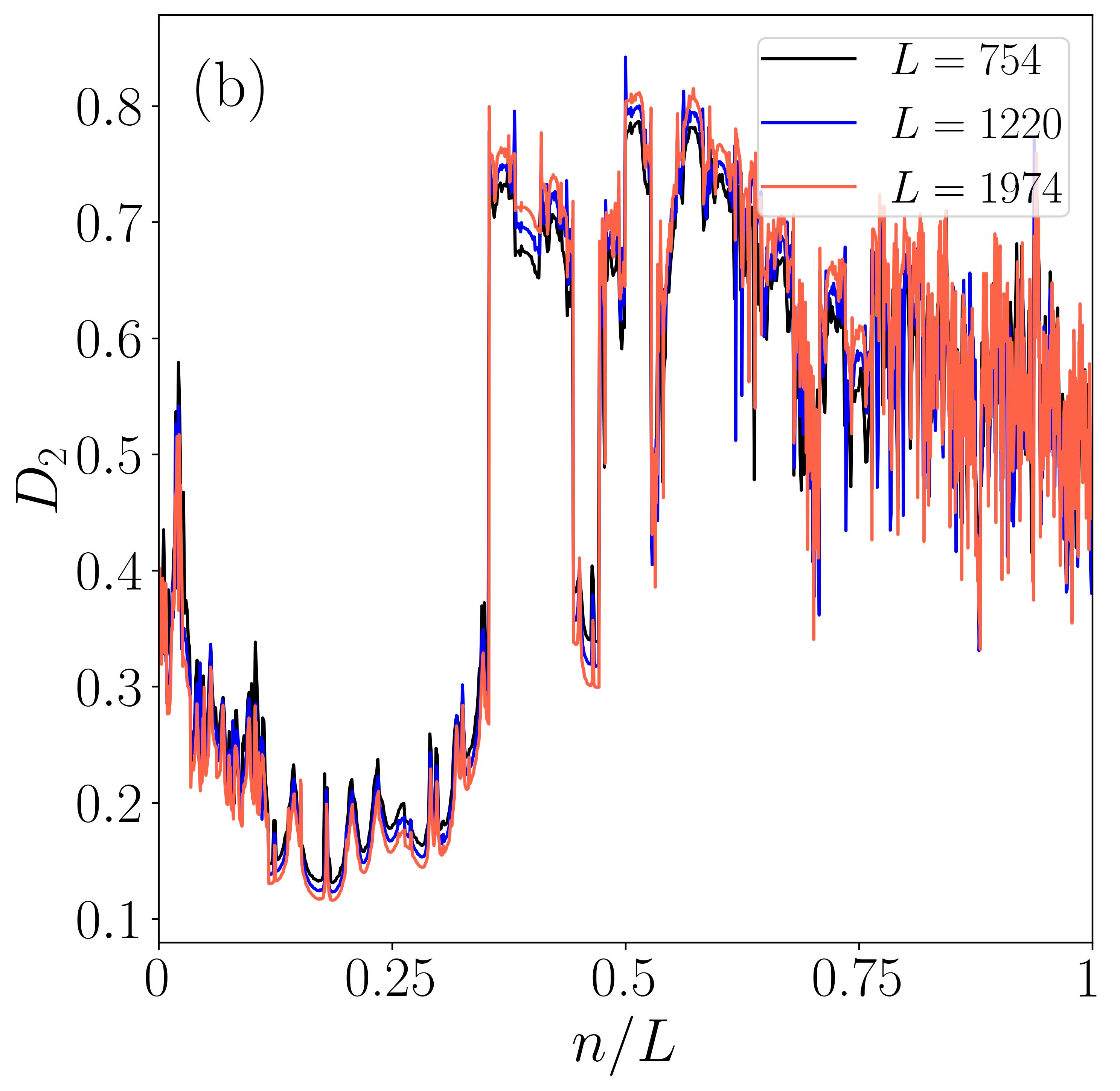}
    \end{subfigure}
    \begin{subfigure}[t]{0.32\textwidth}
        \centering
        \includegraphics[width=\textwidth]{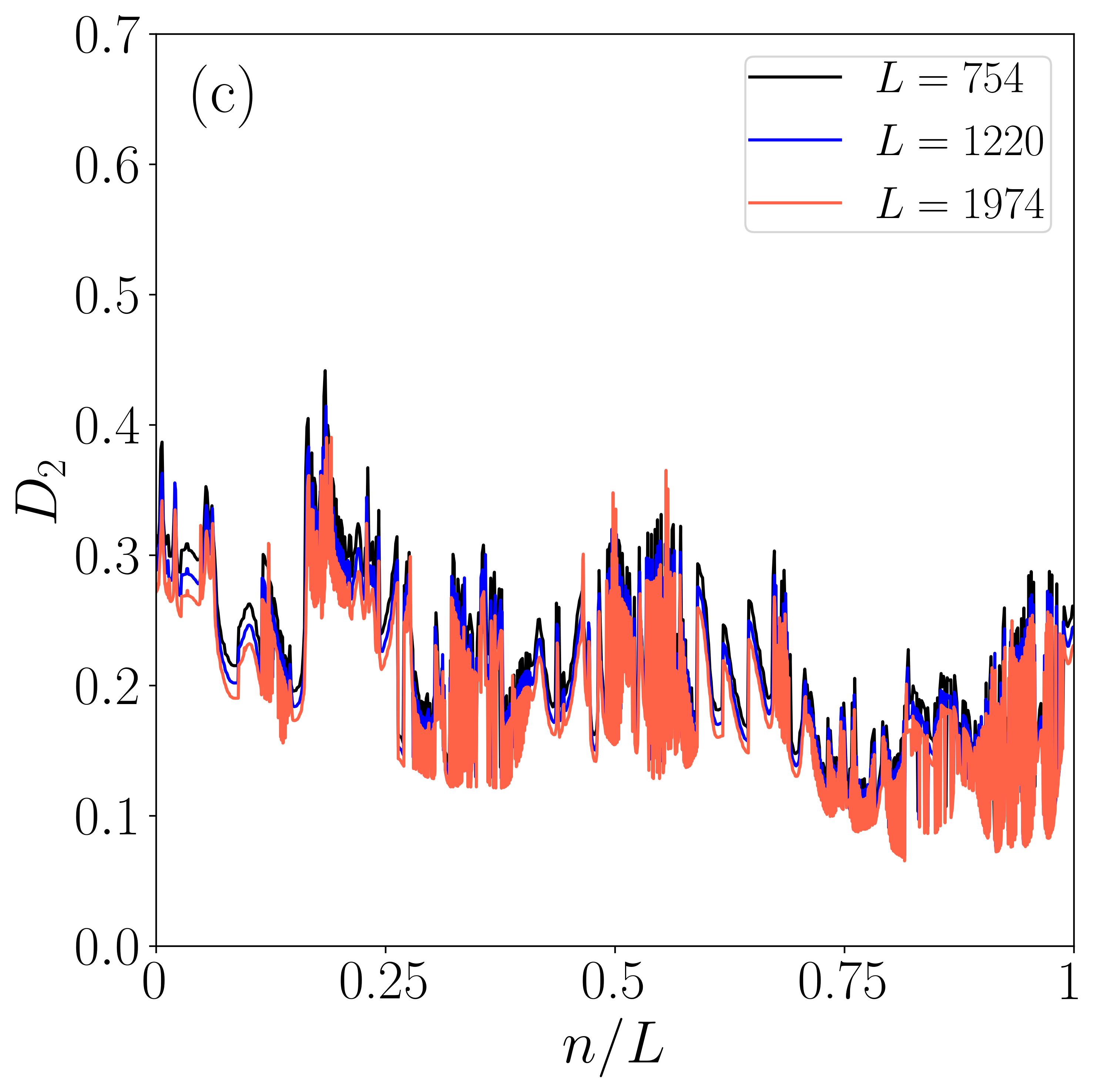}
    \end{subfigure}
    \caption{The fractal dimension $D_{2}$ is plotted as function of $n/L$ corresponding to (a) $\lambda=0.5$, (b) $\lambda=1.5$ and (c) $\lambda=3.5$. The dimerization strength is taken as $\delta=0.5$. The system sizes chosen for this calculation are $L = 754,~1220$, and $1974$.}
    \label{Figure_11}
\end{figure}
\par In Figs.~\ref{Figure_10}(a) and (b), we have shown the phase diagram using $\eta$ value as a function of $\delta$ and $\lambda$ corresponding to the non-driven and the driven cases with the same frequency $\omega=2.5$ for the sake of completeness. The phase diagram contains overall information on the localized, intermediate, and extended phases that appear in the model. By comparing the driven system with the undriven one, it is clear that the effective potential strength decreases in the latter situation. The plot indicates a broader region for the intermediate phase (see Fig.~\ref{Figure_10}(b)). 
\par Furthermore, it is important to note that while $\eta$ gives a global picture differentiating distinct phases, it can not uniquely separate distinct states. As a result, a finite-size analysis is required (shown in Figs.~\ref{Figure_11}(a,b,c)) the quantity, namely, the fractal dimension, $D_2$, which helps in identifying different states accurately and is defined as \cite{genaa4},
\begin{equation}
    D_2= - \lim_{L \rightarrow \infty} \frac{\text{log}(\text{IPR})}{\text{log}(L)}.
    \label{Eqn:fractal}
\end{equation}
It has a value of 1(0) corresponding to the extended(localized) states, in the thermodynamic limit. The critical (multifractal) states have a value between 0 and 1. In Fig.~\ref{Figure_11}(a), we plot $D_{2}$ as a function of the ratio of eigenstate index ($n$) and system size ($L$) corresponding to $\lambda=0.5$. The $D_{2}$ values move towards a value $1$ as the system size is increased, thereby signaling the onset of a completely extended bulk state. \textcolor{red}{On the other hand, in Fig.~\ref{Figure_11}(b) the presence of the multifractal states, in addition to both the localized and the delocalized states in the intermediate phase (corresponding to $\lambda=1.5$) confirm the emergence of multiple mobility edges. For instance, in addition to the mobility edge that separates critical and localized states, there are also mobility edges that distinguish between extended and localized states, as well as those that separate extended and critical states. Recent study \cite{coexist_phases} has also proven that all three phases namely, extended, critical, and localized phases can co-exist in the same parameter space.} In the end, at a higher value of the driving amplitude, namely for $\lambda=3.5$, we observe a complete localization to occur. In all of these three cases, we have chosen $\delta=0.5$ and have considered half of the quasi-energy spectrum owing to the presence of the chiral symmetry.
\par Further, to discern the localization-delocalization transition in the energy spectrum, we chose to look at another numerically accessible quantity that clearly highlights the difference between these distinct phases and hence should widen our analysis of the results. Specifically, we focus on the spectral statistics of the consecutive energy levels \cite{kickedaa2,levelspacingratio1,levelspacingratio2}, which can be captured by the ratio between the adjacent energy gaps, namely, $r_n$ which is defined as,
\begin{equation}
    r_n = \frac{\text{min}\{s_{n},s_{n-1}\}}{\text{max}\{s_{n},s_{n-1}\}}.
    \label{Eqn:level_stat}
\end{equation}
Here, $s_n = E_{n+1}-E_n$, $E_n$ being the $n^{th}$ eigenvalue of the Floquet quasi-energy spectrum. In the localized phase, states that are nearby in energy do not interact or show level repulsion, and as a result nearby energy levels are Poisson distributed with average of $r_n$ over all the eigenvalues have $\langle r_n \rangle \sim 0.39$ \cite{levelspacingratio1}. Whereas, in the extended phase $\langle r_n \rangle$ vanishes.
\par
\begin{figure}[t]
\centering
\includegraphics[width=0.55\linewidth]{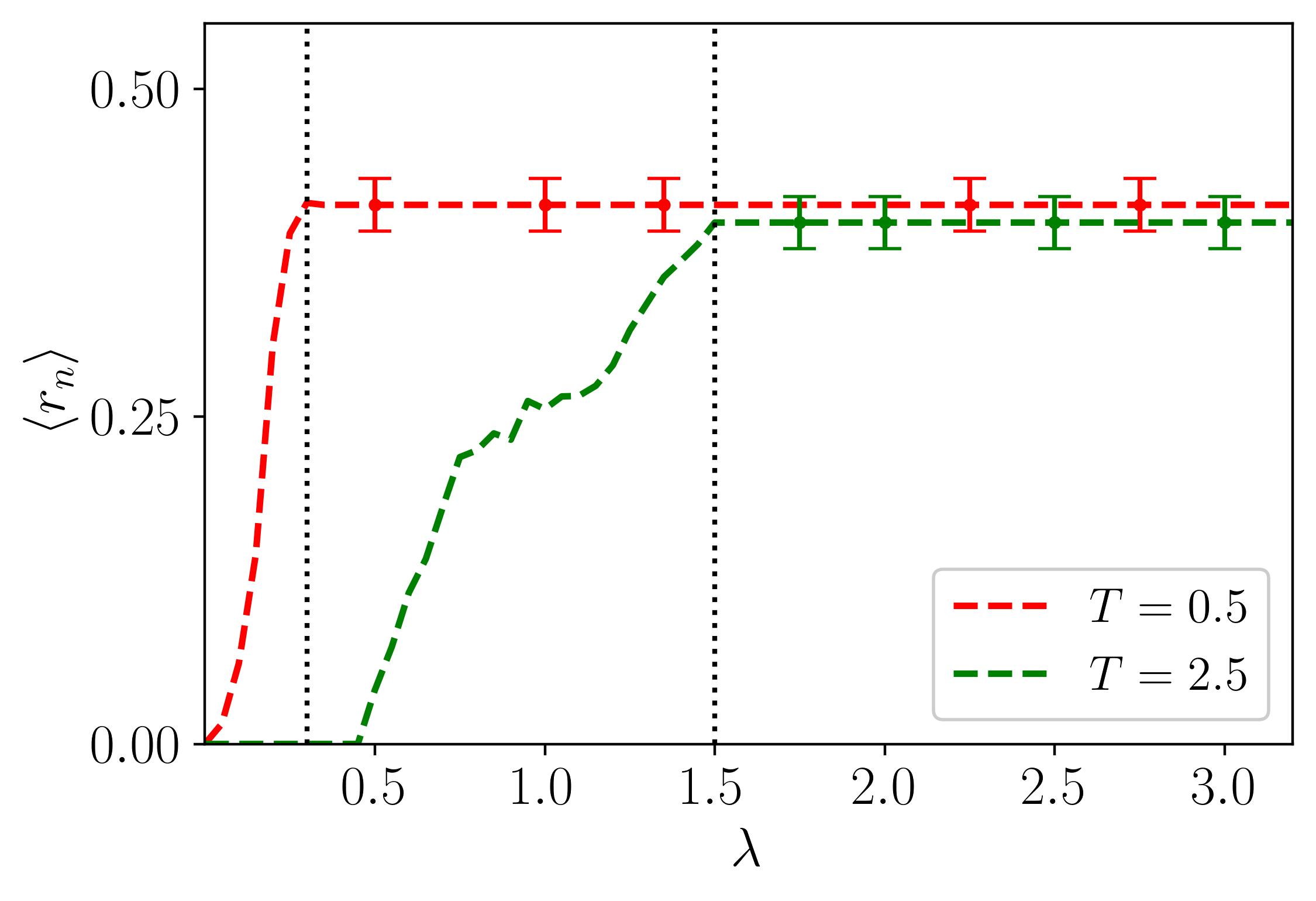}
\caption{The figure shows the variation of $\langle r_n \rangle$ as a function of $\lambda$ for different frequency regimes that are, $T=0.5$ (red dashed line), and $T=2.5$ (green dashed line). the vertical black dotted lines mark the localization transition point. Other parameters are chosen as, $\delta = 0.5$ and $L =1220$.}
\label{Figure_12}
\end{figure}

 In Fig.~\ref{Figure_12}, we have plotted $\langle r_n \rangle$ as a function of disorder strength, $\lambda$ for different frequency regimes. In the high-frequency limit ($T < 1$), there is an abrupt transition to a completely localized phase, indicating that in the high-frequency limit, the system has no extended states. Such a localization behavior at high frequencies validates the effect of the renormalized potential obtained via BCH expansion in Eq.~\ref{BCH_expansion}. While, in the low-frequency limit ($T = 2.5$), such abrupt transition ceases to exist due to the presence of fractal states. For example in the range $0.5 < \lambda < 1.5$, the rise of $\langle r_n \rangle$ is more gradual and hence is an indicative of the presence of critical phase. Moreover, for $\lambda \le 0.5$, $\langle r_n \rangle \approx 0$, indicating the existence of a completely extended phase at low values of disorder.
\par We notice that $\langle r_n \rangle \approx 0$ occurs due to the presence of double degeneracy of the eigenvalues in the extended region. To elucidate this, we introduce the concept of even-odd and odd-even level spacings of the quasi-energy spectrum, which we defined by \cite{odd-evenspacing},
\begin{equation}
   s_{n}^{e-o} = E_{2n} - E_{2n-1}~~;~~s_{n}^{o-e} = E_{2n+1} - E_{2n}.
\end{equation}
\begin{figure}[t]
    \centering
    \begin{subfigure}[t]{0.33\textwidth}
        \centering
        \includegraphics[width=\textwidth]{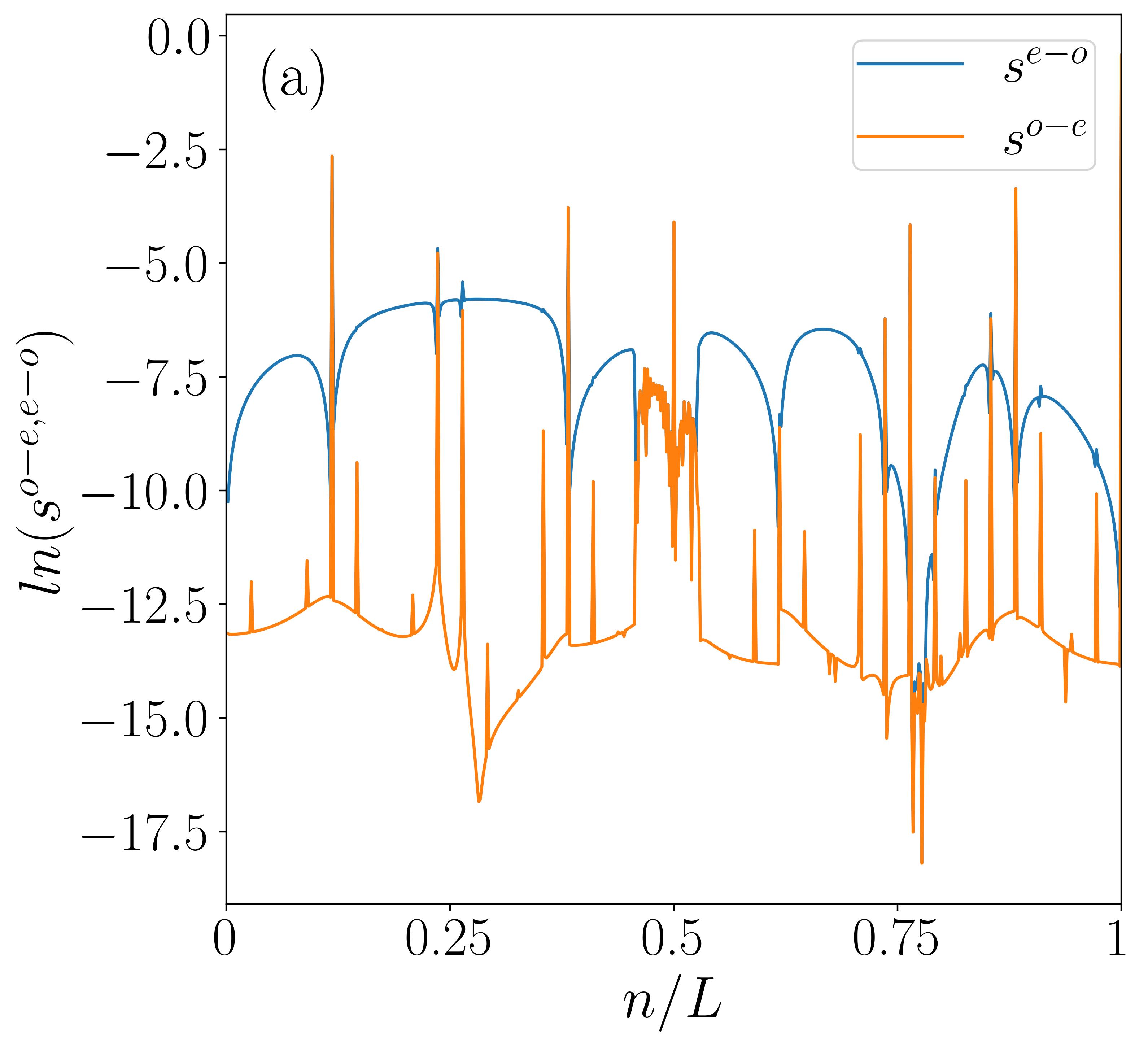}
    \end{subfigure}
    \quad
    \begin{subfigure}[t]{0.32\textwidth}
        \centering
        \includegraphics[width=\textwidth]{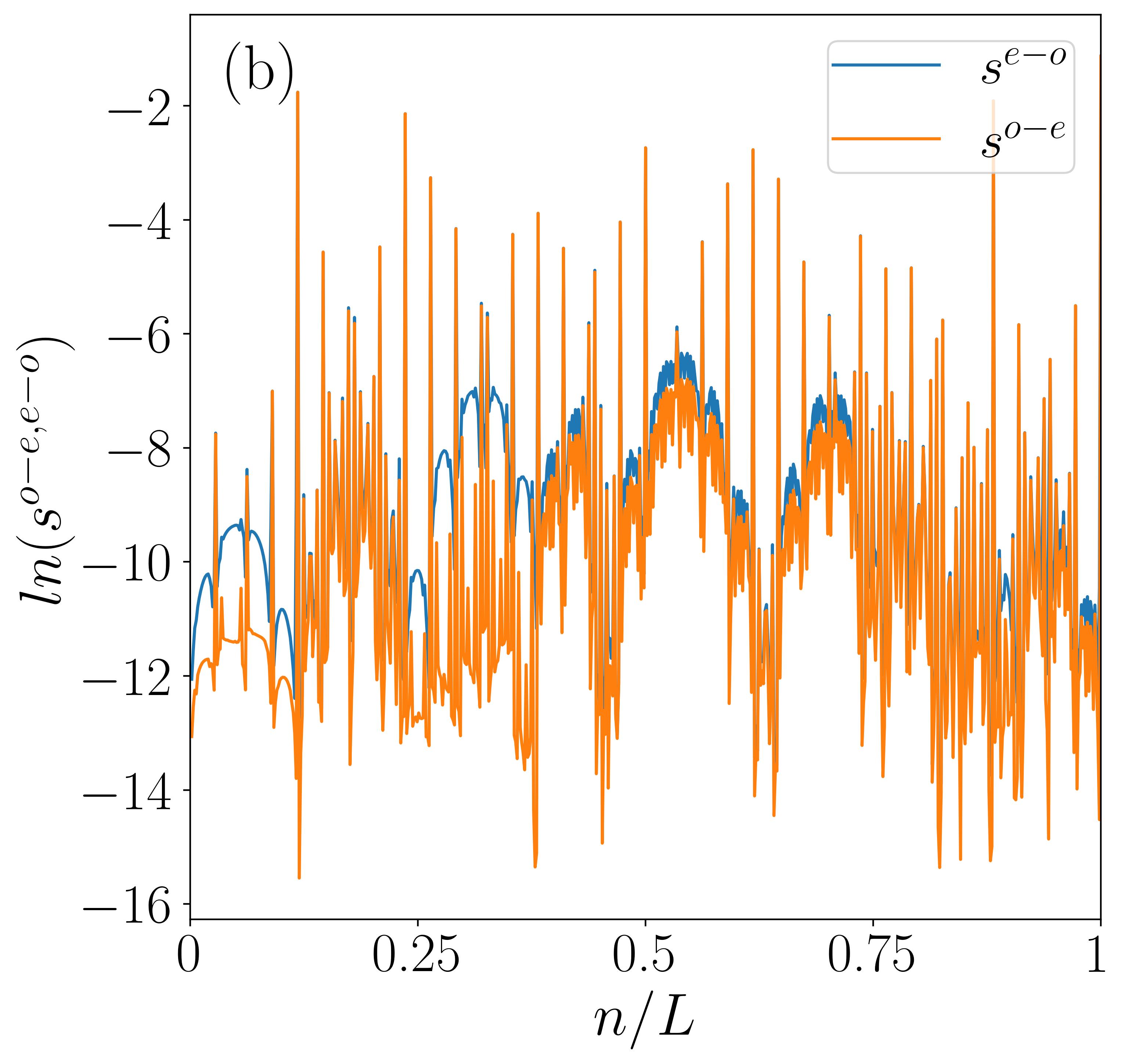}
    \end{subfigure}
    \begin{subfigure}[t]{0.32\textwidth}
        \centering
        \includegraphics[width=\textwidth]{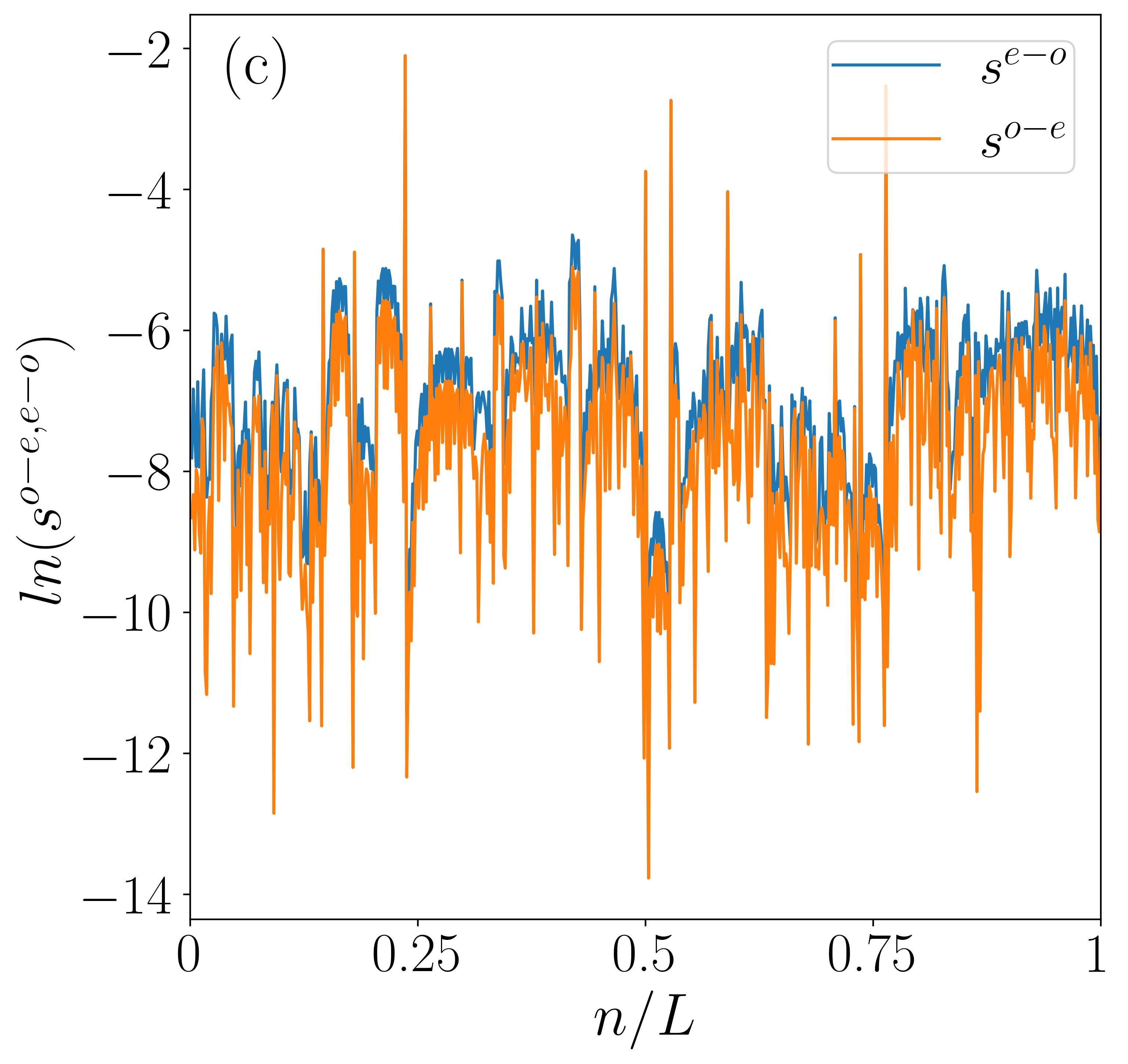}
    \end{subfigure}
    \caption{The even-odd and odd-even level spacings as a function of eigenstate index ($n/L$) are shown in (a): $\lambda=0.5$ (extended phase), (b): $\lambda=1.5$ (intermediate phase), and (c): $\lambda=3.5$ (localized phase) respectively. Other parameters are chosen as, $\omega =2.5$, $\delta = 0.5$ and $L =1220$.}
    \label{Figure_13}
\end{figure}
In Fig.~\ref{Figure_13}, we have plotted the even-odd and odd-even level spacings as a function of site index for a few values
of the disorder, namely, $\lambda = 0.5,1.5,3.5$, corresponding to extended, critical, and the localized phases, respectively. In the extended regime (Fig.~\ref{Figure_13}a), the spectral features exhibit double degeneracy, resulting in a noticeable gap between $s_{n}^{e-o}$ and $s_{n}^{o-e}$. In contrast, within the localized regime (Fig.~\ref{Figure_13}c), both $s_{n}^{e-o}$ and $s_{n}^{o-e}$ show very similar behavior, causing the spectral gap to disappear. Finally, in the intermediate/critical region (Fig.~\ref{Figure_13}b), distributions of $s_{n}^{e-o}$ and $s_{n}^{o-e}$ display significant fluctuations throughout the entire energy spectrum. Moreover, a small gap can be observed between $s_{n}^{e-o}$ and $s_{n}^{o-e}$ at the lower end of the spectrum, indicating that the critical phase can be thought of as an admixture of extended, and localized states.
\par
\begin{figure}[t]
\centering
\includegraphics[width=0.55\linewidth]{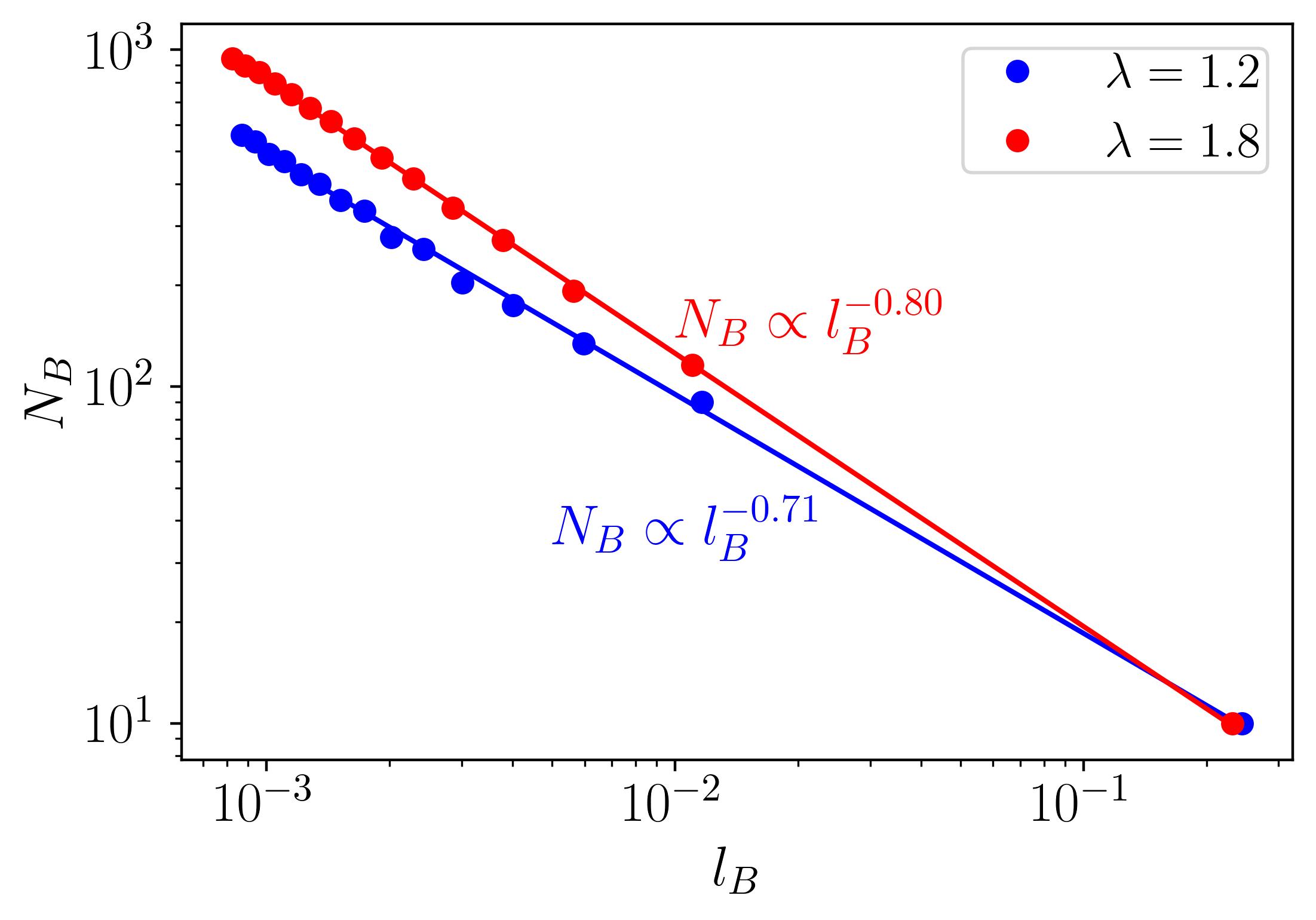}
\caption{The figure shows the variation of $N_B$ as function of $l_B$ in the log-log scale corresponding to $\lambda=1.2$ (blue dots) and $\lambda=1.8$ (red dots). The slopes of these plots give the Hausdorff dimensions, which are obtained as $D_H = 0.71$ and $0.80$ for
$\lambda = 1.2$ and $1.8$, respectively. Other parameters are chosen as, $\omega=2.5$, $\delta = 0.5$ and $L =1220$.}
\label{Figure_14}
\end{figure}

 Further, to unveil the distinct characteristics of the states in the critical regime, we utilize the concept of Hausdorff dimension \cite{hausdorff1,genaa2}. Employing a direct box-counting method for this analysis, we examine the power-law behavior of the total number of boxes (total number of filled states), $N_B$ for a given box length, $l_B$ defined via,
\begin{equation}
    N_B \propto l_{B}^{-D_H}
\label{hausdorff}
\end{equation}
where $D_H$ denotes the Hausdorff dimension corresponding to the energy spectrum. In our case, we compute $D_H$ in two different critical phases corresponding to $\lambda = 1.2$ and $\lambda = 1.8$ respectively. The characteristics of non-trivial fractal can be captured by $0<D_{H}< 1$. In fig.~\ref{Figure_14}, we have plotted $N_B$ as a function of box length $l_B$. We find a linear scaling in the log-log scale, consistent with Eq. (\ref{hausdorff}), and the Hausdorff dimensions are obtained as, $D_H = 0.71$ and $D_H = 0.80$, corresponding to $\lambda=1.2$ and $\lambda=1.8$ respectively. Both the values are notably less than the geometrical dimension, $d=1$. Additionally, we observe a significant difference between the Hausdorff dimension determined in our study and that obtained in previous works at the critical point of the AA model, that is $D_H \sim 0.5$ \cite{hausdorff3}. This distinction can be attributed to the nature of the eigenstates of the AA model, which is characterized by a sharp transition. Further, we get a variety of values for $D_H$ (we have shown two here) which implies different admixtures of the localized and the extended states present in the critical regime. Also, that leads to higher values of Hausdorff dimension compared to that found in the context of  AA model. Therefore, we infer that the dimerized Kitaev chain exhibits a denser spectrum compared to that of the AA model.
\par Summarizing, the interplay among dimerization ($\delta$), QP potential ($\lambda$), and the drive ($T$) generates distinct phases within our system. To gain a comprehensive understanding of the origin of these phases and their transitions, it is crucial to simultaneously analyze the role of each parameter in the concurrent emergence of topological and localization properties in the system.
To begin with, in the static (undriven) case, the system may host topological properties, accompanied by a delocalization-localization transition induced via a strong QP potential. Further, as the value of dimerization increases from zero ($\delta = 0$) to an extreme dimer limit ($\delta = 1$), the chain breaks up into isolated dimers. Thus, the following physical picture emerges: since the unit cells are \emph{``not talking to each other''}, the QP disorder becomes irrelevant. Hence, as the dimerization becomes large, the localization phenomena (with no topological character) set up with ease, even at lower values of the on-site QP strength.

On the other hand, the inclusion of a periodic drive takes the system one step further by notably expediting the localization phenomena as depicted in Fig.~\ref{Figure_9}. Moreover, the periodic drive renormalizes the QP strength (Eq.~\ref{Eqn:renormalized_potential}), which significantly influences the delocalization-localization transition. For instance, all the states are delocalized corresponding to a large time period, $T$ (low-frequency regime) of the drive, while a complete localization of states occurs corresponding to a small period (high-frequency regime). The findings of the low-frequency regime imply the absence of topological characteristics, which may be attributed to the existence of a fully delocalized phase. As the frequency is increased, the potential gets renormalized, which induces some of the states to emerge as localized states. This yields an intermediate phase (with some localized and others delocalized) that holds the key to the delocalization-localization transition. Consequently, this phenomenon could offer prospects of inducing a transition from a trivial to a topological phase as a function of drive frequency. Finally, at very high frequencies, all the states become localized with no topological character. Hence, it is evident that in the driven scenario, the drive frequency, $\omega$, plays a role akin to dimerization, $\delta$, in the static system. It is also intriguing to observe that in the intermediate frequency range, a critical/intermediate phase emerges, characterized by the complete localization of low-energy states alongside the delocalization of high-energy states (Fig.~\ref{Figure_11}). This indicates the emergence of an energy-dependent boundary between them, referred to as the mobility edge. As $\lambda$ increases, the higher-energy states start getting localized. Additionally, since the $\pi$ energy modes appear at the edges of the Floquet Brillouin zone, it explains why we observe $\pi$ energy modes only when the disorder, $\lambda$, is very high.
\par Moreover, to attain a thorough comprehension of the diverse phases emerging within our system and the impact of different components of the model on these phases, we have provided a table (Table.~\ref{table}) that distinctly delineates the various topological and localization characteristics in small and large $\lambda, T$, and $\delta$ regimes.
\par Further, a comparison between our model and several others that exist in the literature will add insights on the roles played by different parameters that derive our model Hamiltonian. To be specific, our system comprises of the following key elements, namely, periodic driving ($T$), dimerization ($\delta$), on-site quasiperiodic potential ($\lambda$), and the $p$-wave superconducting pairing term ($\Delta$). By carefully selecting these parameters, the Hamiltonian can be effectively mapped onto several established models, yielding significant and noteworthy physically realizable outcomes. The associated models under specific conditions and their relevant references are listed below.
\begin{itemize}
    \item The pure Aubry-Andr\'e (AA) model $(\lambda_A = \lambda_B = \lambda$, $\Delta=0$, and $\delta=0$). \cite{aamodel2}
    \item The SSH model ($\lambda=0$, and $\Delta=0$). \cite{ssh}
    \item Disordered SSH model ($\lambda_A = -\lambda_B = \lambda$, and $\Delta=0$). \cite{genaa1}
    \item The Kitaev chain ($\lambda=0$, and $\delta=0$). \cite{kitaev}
    \item The dimerized Kitaev chain ($ \lambda=0$). \cite{dimerizedkitaev1}
\end{itemize}
We commence our exploration with the pure 1D Aubry-Andr\'e (AA) model renowned for its manifestation of a delocalization-localization transition at the critical transition point, $\lambda = 2t$ \cite{aamodel2}.
However, upon introducing $p$-wave superconducting pairing ($\Delta$) to the AA model, the scenario is significantly altered. Notably, the critical transition shifts to a new value, $\lambda = 2(t + \Delta)$, implying that achieving the localization transition necessitates a higher value of $\lambda$, thereby delaying the transition. \cite{disorderedkitaev}. On the other hand, when we only have a dimerization ($\delta$) parameter, the Hamiltonian reduces to the celebrated Su-Schrieffer-Heeger (SSH) model, known for its topological-trivial phase transition occurring at a critical dimerization strength. Further, upon introducing quasiperiodic on-site disorder with the initial condition being $\lambda_A = -\lambda_B = \lambda$ yields a recently discovered phenomenon, namely, the re-entrant localization transition \cite{genaa1}. Now, if we take a re-look at the model under consideration, $\delta=0$ implies a tight binding chain while, $\delta=1$ breaks the lattice into isolated dimers resulting in a completely localized phase. Thus, despite the existence of non-zero pairing ($\Delta$), the localization process, which initially required a higher $\lambda$ in the absence of dimerization in the model \cite{disorderedkitaev}, is now expedited with the inclusion of $\delta$, and even manifests at weak values of $\lambda$. Finally, the incorporation of time-periodic modulation manifests that the driving frequency plays a role similar to the dimerization in the static case.

\begin{table}[!h]
\centering
\hspace{0.01\textwidth}
\begin{minipage}{0.47\textwidth}
\begin{center}
\setlength{\tabcolsep}{10pt} 
\renewcommand{\arraystretch}{3.2}
\begin{tabular}{|| c | c | c  ||}
\hline
 \multicolumn{3}{|c|}{ Static system} \\
\hline
\hline
&Strong $\lambda$&Weak $\lambda$\\
\hline
Strong $\delta$&\shortstack{Trivial and \\ localized phase}& \shortstack{Trivial and \\ localized phase}\\
\hline
Weak $\delta$&\shortstack{Trivial and \\ localized phase}& \shortstack{Topological and \\ extended phase}\\
\hline
\end{tabular}
\end{center}
\end{minipage}%
\hspace{0.005\textwidth} 
\begin{minipage}{0.47\textwidth}
\begin{center}
\setlength{\tabcolsep}{10pt} 
\renewcommand{\arraystretch}{3.2}
\begin{tabular}{|| c | c | c  ||}
\hline
 \multicolumn{3}{|c|}{ Driven system} \\
\hline
\hline
&Strong $\lambda$&Weak $\lambda$\\
\hline
High $T$&\shortstack{Trivial and\\ extended phase}& \shortstack{Trivial and\\ extended phase} \\
\hline
Low $T$&\shortstack{Trivial and\\ localized phase}& \shortstack{Topological and\\ localized phase}\\
\hline
\end{tabular}
\end{center}
\end{minipage}
\caption{{Table depicting various topological and localization phases corresponding to static (left) and dynamic (right) version of the model. Note that strong(weak) $\delta$ refers to $\delta \sim 1 (0)$. As 
$\delta$ increases, the span of zero-energy modes decreases with respect to 
$\lambda$. In the extreme dimer limit ($\delta = 1$), all zero-energy modes vanish, and all bulk states become localized. Further, in a static system, the critical phase is only present within the range $0 < \delta <1$, where the system undergoes various delocalization-localization transitions. Whereas, for a driven system the critical phase only emerges within the intermediate frequency regime. Additionally, it is noteworthy that in the driven system, the frequency, $\omega$ plays a role akin to that of dimerization, $\delta$ in the static system.}}
\label{table} 
\end{table}

\section{\label{conclusions}Conclusions}
In this paper, we consider a one-dimensional dimerized Kitaev chain under a periodic drive of quasiperiodically modulated onsite chemical potential. We have analyzed the topological and localization properties due to the interplay of the periodic drive and the dimerization term present in the system. Based on the behavior of Majorana zero and $\pi$ modes, we have found that low driving frequency induces intriguing observations as compared to the static (undriven) counterpart. Corresponding to the Majorana zero mode, driving induces a phase transition from a trivial to a topologically non-trivial phase, followed by another transition from topological to an Anderson localized phase, which is found for a specific range of the dimerization strength. Most interestingly, a phase, namely the Floquet topological Anderson phase, that consists of the localized $\pi$-modes at the edges associated with completely localized bulk states is found at large values of the driving strength. Further, the localization properties of the bulk states are analyzed. The observation indicates a fully extended phase corresponding to a low-frequency range, while a complete localization of all the states is established at the high-frequency regime. However, localized, critical, and extended phases co-exist at an intermediate frequency region. Further, we have established our results via a finite-size scaling analysis of the fractal dimension. Also, we have looked at the mean level spacing corresponding to low and high-frequency regimes, where distinct features are noted for the localized and the critical phases. Finally, the study of Hausdorff dimension elucidates distinct attributes unique to the critical phase, providing deeper insights into its properties.
\par Our insights that a static system in presence of QP potential simulates rich delocalization-localization phenomenon under non-equilibrium settings allow us to suggest a route for the experimental validation of our
results. Interestingly, owing to the intrinsic simplicity of circuit design, significant advancements have been made in topolectrics, affirming the presence of Majorana modes \cite{topoelectric1,topoelectric2,topoelectric3}. Further, a dimerized Kitaev chain can be constructed by considering two arrays of inductors within an LC network such that each sublattice is simulated by an inductor \cite{kitaeveelectric}. The top array can then be contemplated as particles and the bottom array as antiparticles. 
Next, the inclusion of periodic drive as outlined in Ref. \cite{floquetelectric}, entails coupling each harmonic of the drive (also termed as Floquet replicas), to the primary circuit via capacitive connections, that needs to match the strength of the drive. Finally, to analyze the circuit's localization behavior, one may opt for the simplest experimentally measurable parameter, the two-port impedance factor as illustrated in Ref. \cite{topoelectric3}.

\section*{Data availability statement}
All data and codes that support the findings of this study can be obtained from the corresponding author upon request.

\section*{Author contributions}
All the authors have contributed equally to the manuscript.

\end{document}